\documentclass[aps,prl,preprint,superscriptaddress]{revtex4-1}

\usepackage{graphicx}
\usepackage{multirow}
\usepackage{amsmath}
\usepackage{bm}
\usepackage{setspace}
\usepackage{xcolor}

\begin{document}

\title{Coulomb explosion imaging of concurrent CH$_2$BrI photodissociation dynamics}

\author{Michael Burt}
\affiliation{The Chemistry Research Laboratory, Department of Chemistry, University of Oxford, Oxford OX1 3TA, United Kingdom}
\author{Rebecca Boll}
\affiliation{Deutsches Elektronen-Synchrotron DESY, Notkestra\ss e 85, 22607 Hamburg, Germany}
\author{Jason W. L. Lee}
\author{Kasra Amini}
\author{Hansjochen K\"ockert}
\author{Claire Vallance}
\affiliation{The Chemistry Research Laboratory, Department of Chemistry, University of Oxford, Oxford OX1 3TA, United Kingdom}
\author{Alexander S. Gentleman}
\author{Stuart R. Mackenzie}
\affiliation{The Physical and Theoretical Chemistry Laboratory, Department of Chemistry, University of Oxford, Oxford OX1 3QZ, United Kingdom}
\author{Sadia Bari}
\author{C\'edric Bomme}
\author{Stefan D\"usterer}
\author{Benjamin Erk}
\author{Bastian Manschwetus}
\author{Erland M\"uller}
\author{Dimitrios Rompotis}
\author{Evgeny Savelyev}
\author{Nora Schirmel}
\affiliation{Deutsches Elektronen-Synchrotron DESY, Notkestra\ss e 85, 22607 Hamburg, Germany}
\author{Simone Techert}
\affiliation{Deutsches Elektronen-Synchrotron DESY, Notkestra\ss e 85, 22607 Hamburg, Germany}
\affiliation{Max Planck Institute for Biophysical Chemistry, 37077 G\"ottingen, Germany}
\affiliation{Institute of X-ray Physics, University of G\"ottingen, 37077 G\"ottingen, Germany}
\author{Rolf Treusch}
\affiliation{Deutsches Elektronen-Synchrotron DESY, Notkestra\ss e 85, 22607 Hamburg, Germany}
\author{Jochen K\"upper}
\affiliation{Center for Free-Electron Laser Science, Deutsches Elektronen-Synchrotron DESY, Notkestra\ss e 85, 22607 Hamburg, Germany}
\affiliation{Center for Ultrafast Imaging, Universit\"at Hamburg, Luruper Chaussee 149, 22761 Hamburg, Germany}
\affiliation{Department of Physics, Universit\"at Hamburg, Luruper Chaussee 149, 22761 Hamburg, Germany}
\author{Sebastian Trippel}
\affiliation{Center for Free-Electron Laser Science, Deutsches Elektronen-Synchrotron DESY, Notkestra\ss e 85, 22607 Hamburg, Germany}
\affiliation{Center for Ultrafast Imaging, Universit\"at Hamburg, Luruper Chaussee 149, 22761 Hamburg, Germany}
\author{Joss Wiese}
\affiliation{Center for Free-Electron Laser Science, Deutsches Elektronen-Synchrotron DESY, Notkestra\ss e 85, 22607 Hamburg, Germany}
\author{Henrik Stapelfeldt}
\affiliation{Department of Chemistry, Aarhus University, Langelandsgade 140, DK-8000 Aarhus C, Denmark}
\author{Barbara Cunha de Miranda}
\author{Renaud Guillemin}
\author{Iyas Ismail}
\author{Lo\"ic Journel}
\author{Tatiana Marchenko}
\author{J\'er\^ome Palaudoux}
\author{Francis Penent}
\affiliation{Sorbonne Universit\'es, UPMC Universit\'e Paris 06, CNRS, UMR 7614, Laboratoire de Chimie Physique-Mati\`ere et Rayonnement, F-75005 Paris, France}
\author{Maria Novella Piancastelli}
\affiliation{Sorbonne Universit\'es, UPMC Universit\'e Paris 06, CNRS, UMR 7614, Laboratoire de Chimie Physique-Mati\`ere et Rayonnement, F-75005 Paris, France}
\affiliation{Department of Physics and Astronomy, Uppsala University, PO Box 516, 75120 Uppsala, Sweden}
\author{Marc Simon}
\author{Oksana Travnikova}
\affiliation{Sorbonne Universit\'es, UPMC Universit\'e Paris 06, CNRS, UMR 7614, Laboratoire de Chimie Physique-Mati\`ere et Rayonnement, F-75005 Paris, France}
\author{Felix Brausse}
\affiliation{Max-Born-Institut, 12489 Berlin, Germany}
\author{Gildas Goldsztejn}
\affiliation{Sorbonne Universit\'es, UPMC Universit\'e Paris 06, CNRS, UMR 7614, Laboratoire de Chimie Physique-Mati\`ere et Rayonnement, F-75005 Paris, France}
\affiliation{Max-Born-Institut, 12489 Berlin, Germany}
\author{Arnaud Rouz\'ee}
\affiliation{Max-Born-Institut, 12489 Berlin, Germany}
\author{Marie G\'el\'eoc}
\author{Romain Geneaux}
\author{Thierry Ruchon}
\affiliation{LIDYL, CEA, CNRS, Universit\'e Paris-Saclay, CEA-Saclay 91191 Gif-sur-Yvette, France}
\author{Jonathan Underwood}
\affiliation{Department of Physics and Astronomy, University College London, London WC1E 6BT, United
Kingdom}
\author{David M. P. Holland}
\affiliation{Daresbury Laboratory, Daresbury, Warrington, Cheshire WA4 4AD, United Kingdom}
\author{Andrey S. Mereshchenko}
\author{Pavel K. Olshin}
\affiliation{Saint-Petersburg State University, 7/9 Universitetskaya nab., St. Petersburg, 199034 Russia}
\author{Per Johnsson}
\author{Sylvain Maclot}
\author{Jan Lahl}
\affiliation{Department of Physics, Lund University, 22100 Lund, Sweden}
\author{Artem Rudenko}
\author{Farzaneh Ziaee}
\affiliation{J. R. Macdonald Laboratory, Department of Physics, Kansas State University, Manhattan, Kansas 66506, USA}
\author{Mark Brouard} \email{mark.brouard@chem.ox.ac.uk}
\affiliation{The Chemistry Research Laboratory, Department of Chemistry, University of Oxford, Oxford OX1 3TA, United Kingdom}
\author{Daniel Rolles}
\affiliation{J. R. Macdonald Laboratory, Department of Physics, Kansas State University, Manhattan, Kansas 66506, USA}

\date{\today}

\textcolor{red}{}
\begin{abstract}
The dynamics following laser-induced molecular photodissociation of gas-phase CH$_2$BrI at 271.6\,nm were investigated by time-resolved Coulomb explosion imaging using intense near-IR femtosecond laser pulses. The observed delay-dependent photofragment momenta reveal that CH$_2$BrI undergoes C-I cleavage, depositing 65.6\% of the available energy into internal product states, and that absorption of a second UV photon breaks the C-Br bond of CH$_2$Br. Simulations confirm that this mechanism is consistent with previous data recorded at 248\,nm, demonstrating the sensitivity of Coulomb explosion imaging as a real-time probe of chemical dynamics.
\end{abstract}

\pacs{33.80.Gj, 33.15.Hp, 33.80.Rv, 34.50.Gb}

\maketitle


\section{I. Introduction}
Femtosecond optical pulses allow chemical reactions to be studied on their natural time scales \cite{Zewail1988, Dura2008, DeNalda2008, Corrales2014, Ibrahim2014}. At pulse intensities of 10$^{14}$--10$^{16}$\,W\,cm$^{-2}$, several electrons can be detached from an isolated molecule through tunnel or multiphoton ionization, creating an unstable cation that Coulomb explodes into fragments. If the explosion occurs before the molecule rearranges, its original geometry can be established from the fragment momenta \cite{Vager1989a, Cornaggia1992, Hering1999, Sanderson1999, Legare2005, Gagnon2008, Corrales2012, Pitzer2013, Slater2014, Christensen2015, Christiansen2016, Frasinski1989, Frasinski2016}. This has been used to determine the structures and chirality of gas-phase molecules \cite{Kitamura2001,Pitzer2013,Pickering2016}. Time-resolved dynamics can be measured by coupling Coulomb explosion imaging with femtosecond pump-probe spectroscopy \cite{Stapelfeldt1995, Ergler2005a, Legare2006b, Christensen2014, Erk2014, Ibrahim2014, Boll2016}.

The work described in this article applies timed Coulomb explosion imaging to a fundamental process, UV-induced photochemistry, by probing the concurrent dissociation pathways of CH$_2$BrI at 271.6\,nm. We demonstrate how dynamical properties, such as the extent of internal energy deposition, can be gauged from time-dependent data. Studies of molecular photodissociation dynamics are generally limited to simpler molecular systems, and are often reliant on spectroscopic probes of one or more of the photofragments. Because the Coulomb explosion process is a more universal detection method than conventional spectroscopic techniques, such as resonantly enhanced multiphoton ionization, Coulomb explosion imaging opens up the possibility of studying photodissociation dynamics in a much wider range of molecular systems. The present work provides initial proof-of-principle for this experimental strategy.

Dihalomethanes offer a rich source of photochemistry in the gas- and solution-phases \cite{Lee1982, Butler1987, Xu2002, Schmitt1987, Senapati2002, Senapati2004, Zhang2005, Cheng2016, El-Khoury2010, Tang2010}. Their geminal halogens allow a complex system of repulsive excited states to be accessed in the UV, making them archetypes for mode-selective chemistry. The CH$_2$BrI absorption spectrum exhibits three short-wave UV transitions: two are centered at 268\,nm and 213\,nm and respectively arise from promotion of non-bonding iodine or bromine electrons to antibonding C-I and C-Br orbitals, while the third at 190\,nm corresponds to iodine Rydberg {\em ns} $\rightarrow$ {\em np} transitions \cite{Lee1982, Butler1987}. These features are sufficiently distinct in energy to allow specific photodissociation pathways to be selected. Photoexcitation at 271.6\,nm is near the maximum of the {\em n}(I) $\rightarrow$ $\sigma$$^{*}$(C-I) transition and preferentially breaks the C-I bond. Comparatively, the proportion of C-I and C-Br bond cleavages at 248\,nm is 6:5, and decreasing the wavelength to 193\,nm or 210\,nm promotes both C-Br cleavage and, following isomerization to {\em iso}-CH$_2$BrI, IBr elimination \cite{Butler1987}. Similar behaviour has been reported for CH$_2$ClI \cite{Schmitt1987, Senapati2002, Senapati2004, Zhang2005, Cheng2016}.

The energetics of dihalomethanes allow primary or secondary dissociation reactions to be accurately distinguished through the measured momenta of the observed products. For CH$_2$BrI, a single 271.6\,nm (4.57\,eV) photon is sufficient to overcome the C-I and C-Br dissociation thresholds at 2.39\,eV and 2.94\,eV, and to induce IBr elimination (3.84\,eV). Secondary or three-body dissociations, by contrast, require two UV photons to create CH$_2$, I, and Br (5.66\,eV) \cite{Butler1987, Rosenstock1977, Kudchadker1978}. Fragment momenta measured following neutral photodissociation will reflect these thresholds and the energy deposited into internal photofragment states. They can therefore be correlated using Coulomb explosion imaging to determine the photodissociation mechanism of the parent molecule \cite{Butler1987, Gagnon2008}.

\section{II. Methods}
CH$_2$BrI fragment momenta were recorded using the CFEL-ASG MultiPurpose (CAMP) instrument at the free-electron laser in Hamburg (FLASH, beamline BL1) \cite{Struder2010, Erk2017}. CAMP houses a double-sided time-of-flight mass spectrometer tuned for simultaneous velocity map imaging (VMI) of ions and electrons. CH$_2$BrI was expanded into CAMP through two skimmers using a continuous jet. The collimated and neat molecular beam was then intersected in the VMI interaction region by the frequency tripled output ($\lambda_{pump}$ = 271.6\,nm, bandwidth = 2.5\,nm (FWHM)) and fundamental output ($\lambda_{probe}$ = 815\,nm, bandwidth = 26\,nm (FWHM)) of the FLASH pump-probe laser, which uses a 10\,Hz Ti:sapphire oscillator and chirped pulse amplifier to produce 55\,fs IR pulses \cite{Redlin2011}. UV pulses were created by splitting the IR beam and sequentially passing it through a 50\,$\mu$m thick tripling crystal on a 1\,mm substrate and a prism compressor. Typical IR and UV pulse energies were 620\,$\mu$J and 22.5\,$\mu$J, and these were linearly polarized parallel to the detector and focused to a $40 \times 60\,\mu$m$^{2}$ spot within the interaction region. The time between pulses was varied in 50\,fs steps over three picoseconds ($-0.35$\,ps to 2.65\,ps) using an automated delay stage. Negative and positive delays respectively correspond to the IR and UV pulses arriving first. The temporal overlap of the two pulses, $t_{\rm 0}$, was determined from the cross-correlation intensity maxima of the I$^{+}$ and CH$_2$Br$^{+}$ fragments, the standard deviation was 72\,fs, yielding a FWHM time resolution of 170\,fs.

Under velocity mapping conditions \cite{Eppink1997}, fragments generated in the CAMP chamber with the same mass and velocity are focused to the same point on a two-dimensional detector array comprising two chevron-stacked microchannel plates and a P47 phosphor screen. As these ions are additionally separated by their time-of-flight to the detector, the result is a series of images for each mass-to-charge ratio ($m/z$) where the Abel inverted image radii correspond to the ion momenta.  Time-resolved dynamics are investigated by acquiring ion images at discrete delay steps and tracking changes in the radial intensity distributions.

Photons emitted by the phosphor were recorded by a Pixel Imaging Mass Spectrometry (PImMS) camera employing the $324 \times 324$ pixel PImMS2 sensor. PImMS2 is an event-triggered device that logs the position ($x,y$) and arrival time ($t$, relative to an external trigger) of incident photons to a precision of 40\,ns, allowing every mass-resolved fragment ion to be imaged and correlated within one experiment \cite{Clark2012, John2012}.  Fast imaging sensors have several advantages over conventional CCD cameras. The latter allow imaging of single fragments, or small time windows of fragments, and consequently miss correlations between ions with different $m/z$. By contrast, the PImMS camera can be used for coincident and covariance imaging \cite{Slater2015}. In the present experiment, the camera repetition rate was synchronized to the 10\,Hz pump-probe laser system, and approximately 450 laser shots were taken per delay step. It should be noted that ions impacting the detector often lead to phosphor flashes bright enough to activate more than one PImMS pixel. These pixel clusters were centroided to improve the resolution of the ion images \cite{Slater2014}.

\section{III. Results}
\begin{figure}
\includegraphics{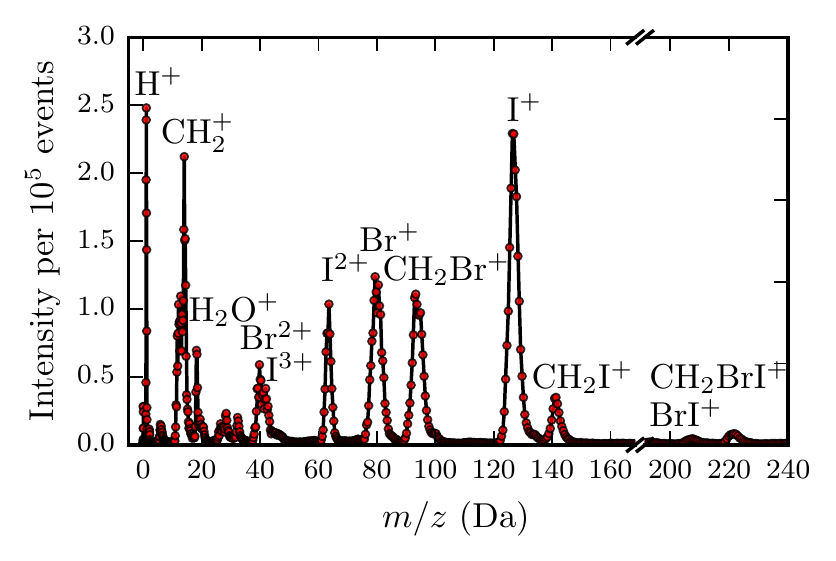}
\caption{The centroided CH$_2$BrI time-of-flight spectrum accumulated by PImMS during the 0\,ps to 2.65\,ps UV-IR delay scan. Each red circle represents a time stamp and an ion image. \label{fig1}}
\end{figure}

\begin{figure}
\includegraphics{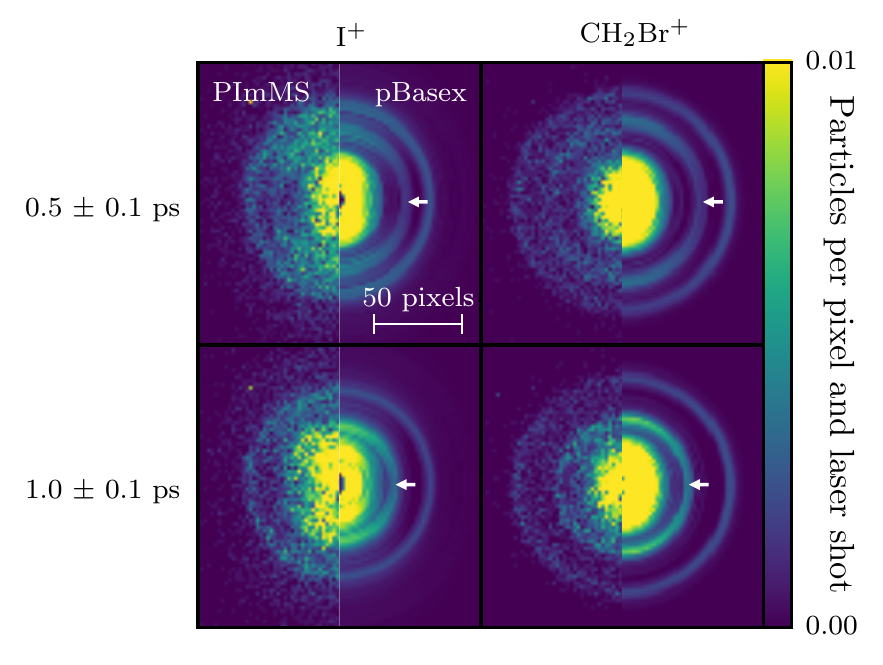}
\caption{I$^{+}$ and CH$_2$Br$^{+}$ PImMS images before and after pBasex Abel inversion acquired at pump-probe delays of $0.50 \pm 0.10$ and $1.00 \pm 0.10$\,ps. The arrowed rings contract with increasing delay, and are consistent with a Coulomb explosion occurring after neutral photodissociation. \label{fig2}}
\end{figure}

Figures\ \ref{fig1} and \ref{fig2} illustrate a PImMS time-of-flight spectrum and its corresponding velocity map ion images. The mass spectrum was accumulated from {\em ca.}\ 25,000 laser shots acquired at positive delays. Each point represents ($x,y,t$) data integrated over $x$ and $y$ for a particular time stamp. Figure\ \ref{fig2} presents images generated by integrating the same data over the I$^{+}$ and CH$_2$Br$^{+}$ time-of-flight peaks at two delays, $0.50 \pm 0.10$\,ps and $1.00 \pm 0.10$\,ps. The images each exhibit two rings that are independent of the pump-probe delay, and one, designated by an arrow, that contracts with increasing delay. Such features are evident in several fragments (I$^{+}$, Br$^{+}$, CH$_2$I$^{+}$, CH$_2$Br$^{+}$, and CH$_2$$^{+}$), and are assigned below to characterize the reaction mechanism. Animations of the delay-dependent I$^{+}$ and CH$_2$Br$^{+}$ images are also provided as Supplemental Material.

\begin{figure*}
\includegraphics{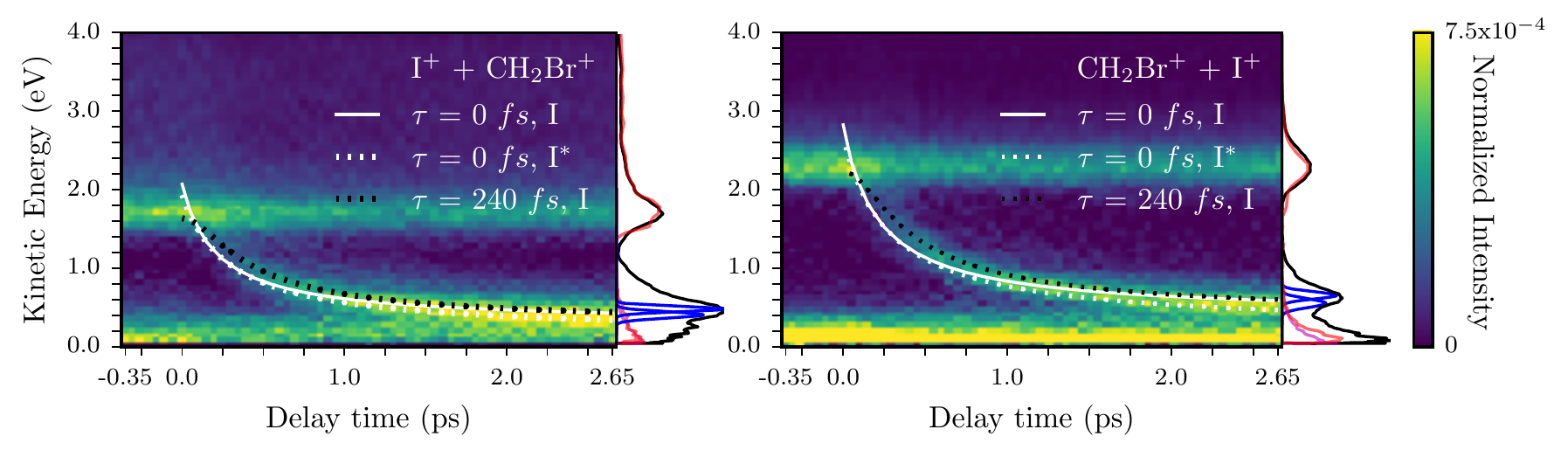}
\caption{I$^{+}$ (left) and CH$_2$Br$^{+}$ (right) kinetic energy distributions, obtained from pBasex Abel inverted ion images, as a function of pump-probe delay. Projections of the 1.85\,ps data are shown to the right of each figure (black lines) and compared with normalized UV (violet lines) and IR background (red lines), as well as the sum of the asymptotic experimental data from Ref. \cite{Butler1987} and the electrostatic kinetic energy at 1.85\,ps (blue lines). Simulations of the expected kinetic energy are overlaid on the figures and labeled as X$^{+}$ + Y$^{+}$ for the plotted fragment X and co-fragment Y. The I({\em $^{2}$P$_{3/2}$}) (I, white solid lines) and  I({\em $^{2}$P$_{1/2}$}) (I$^{*}$, white dotted lines) pathways were modeled by assuming that the neutral fragments instantly reach their final velocities following photodissociation. The black dotted curves model I({\em $^{2}$P$_{3/2}$}) +  CH$_2$Br acceleration using a 240\,fs lifetime. \label{fig3}}
\end{figure*}

The time-dependencies of the ion momenta in Fig.\ \ref{fig2} are illustrated by plotting ion image radial intensities against the pump-probe delay. Figure\ \ref{fig3} presents the kinetic energy distributions of I$^{+}$ and CH$_2$Br$^{+}$ after pBasex Abel inversion of the respective ion images at each delay step \cite{Garcia2004}. Normalized data from 1.85\,ps, the delay step with the most statistics (7,000 laser shots), are also projected to the right of each figure and compared with the UV and IR background. Each distribution contains two invariant channels and one delay-dependent curve that begins when both laser pulses are overlapped at $t_{0}$, and which decays to a limiting value as the probe pulse is delayed. The IR background and the negative delay regions, where the IR light arrives first, both indicate that the I$^{+}$ and CH$_2$Br$^{+}$ channels centered at $1.73 \pm 0.17$\,eV and $2.31 \pm 0.21$\,eV are caused by the double ionization and Coulomb explosion of CH$_2$BrI. These features match their expected electrostatic kinetic energies of 1.76\,eV and 2.41\,eV when the two charges are separated by the equilibrium I-Br distance of CH$_2$BrI, 345\,pm \cite{Liu2005, Kim2014, Bailleux2014b}, and exhibit a decrease in ion yield at positive delays that corresponds to the depletion of CH$_2$BrI by the UV pulse. By contrast, if the charges were located on iodine and carbon, the I$^{+}$ and CH$_2$Br$^{+}$ kinetic energies would be 2.82\,eV and 3.85\,eV, respectively, supporting the view that the CH$_2$Br$^{+}$ charge is located on bromine \cite{Liu2005, Kim2014, Bailleux2014b}. The near-zero kinetic energy channels are attributed to strong-field or UV multiple photon ionization yielding I$^{+}$ or CH$_2$Br$^{+}$ with neutral co-fragments.

The delay-dependent channels illustrate strong-field ionization following UV-induced photodissociation into neutral fragments. As CH$_2$BrI is internally cold prior to the arrival of the UV pulse, the total kinetic energy, $T$, of the ionized photofragments at a given delay is the sum of the available translational energy following photodissociation and the electrostatic (Coulombic) potential energy:
\begin{equation}
T = (h\nu - D_{0} - E_{{\rm int}}^{*} - E_{{\rm so}}) + \frac{k_{\rm e}q_{\rm A}q_{\rm B}}{r_{{\rm AB}}} \,.
\end{equation}
In the above equation, $k_{{\rm e}}$, $q_i$, and $r_{{\rm AB}}$ are the electrostatic constant, the charge on fragment $i$, and distance between the charges on fragments A and B. The latter is evaluated at each delay by assuming the fragments instantly reach their final velocities following two-body dissociation, or by modeling these velocities as functions of time. The distance traveled during the delay period is then added to the equilibrium internuclear separation at $t_{0}$. $D_{0}$ and $E_{{\rm so}}$ are the photodissociation threshold and energy difference between the ground ({\em $^{2}$P$_{3/2}$}) and excited ({\em $^{2}$P$_{1/2}$}) halogen spin-orbit states. As these are known for C-I ($D_{0} = 2.39$\,eV, $E_{{\rm so}} = 0.941$\,eV) and C-Br ($D_{0}  = 2.94$\,eV, $E_{{\rm so}} = 0.457$\,eV) cleavage \cite{Butler1987}, the internal excitation of the products, $E_{{\rm int}}^{*}$, can be determined from the difference between the photon energy at 271.6\,nm, $h\nu$, and the asymptotic value of the total product kinetic energy at long pump-probe delays. In this work, the asymptotic kinetic energies match those reported at 248\,nm (see Section IV), so photodissociation was modeled using established $E_{{\rm int}}^{*}$ values for the I({\em $^{2}$P$_{3/2}$}) and I({\em $^{2}$P$_{1/2}$}) channels (68.1\% and 67.8\% of the available energy following photodissociation, respectively) \cite{Butler1987}.

\begin{figure}
\includegraphics{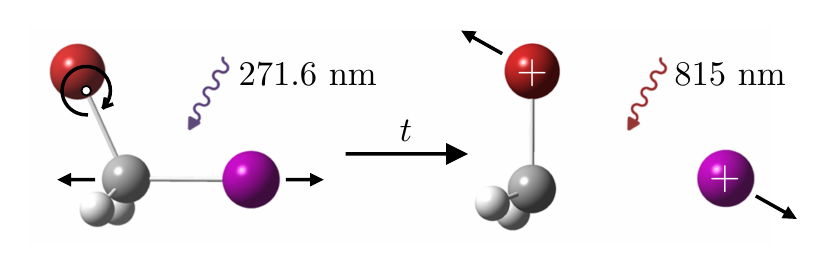}
\caption{Schematic of the photodissociation (left) and Coulomb explosion (right) of CH$_2$BrI. C, H, Br, and I are respectively colored gray, white, red, and purple. Absorption at 271.6\,nm primarily breaks the C-I bond, creating fragments with conserved momenta along the C-I axis. This force rotationally excites CH$_2$Br as its centre of mass (white circle) lies near bromine. After a set delay $t$, strong-field ionization of the co-fragments at 815\,nm results in electrostatic repulsion. \label{fig4}}
\end{figure}

From momentum conservation, the fragment kinetic energies following two-body breakup are related by their mass $m$:
\begin{equation}
T_{\rm A} = T\frac{m_{\rm B}}{m_{{\rm AB}}} \qquad T_{\rm B} = T\frac{m_{\rm A}}{m_{{\rm AB}}} \,.
\end{equation}
The delay-dependent I$^{+}$ and CH$_2$Br$^{+}$ kinetic energy channels in Fig.\ \ref{fig3} exhibit conserved momenta, indicating that they result from C-I cleavage. This process, which is depicted in Fig.\ \ref{fig4}, rotationally excites CH$_2$Br since the product velocities are initially directed along the C-I axis, while the CH$_2$Br centre of mass lies about 30\,pm from bromine along the C-Br bond. The extent of CH$_2$Br excitation is reflected by the I$^{+}$ and CH$_2$Br$^{+}$ asymptotic energies of $0.25 \pm 0.02$\,eV and $0.35 \pm 0.02$\,eV, which are collectively lower than the 2.18\,eV available following UV absorption. These energies include contributions from both the I({\em $^{2}$P$_{3/2}$}) and I({\em $^{2}$P$_{1/2}$}) pathways as the two are not resolved here. However, I({\em $^{2}$P$_{3/2}$}) and I({\em $^{2}$P$_{1/2}$}) were observed in a 4:3 ratio at 248 nm, and this is expected to remain approximately the same at 271.6\,nm based on the behaviour of CH$_2$ClI and CH$_{3}$I \cite{Parker1998, Senapati2002, Senapati2004}. Using this ratio, the asymptotic kinetic energies indicate that CH$_2$Br is internally excited by 65.6 $\pm$ 1.6\% of the available energy, matching the reported {\em E$_{{\rm int}}^{*}$} value of 67.9 $\pm$ 1.5\% \cite{Butler1987}. At 271.6 nm, the latter proportion yields asymptotic I$^{+}$ and CH$_2$Br$^{+}$ kinetic energies of 0.29 and 0.40 eV for the I({\em $^{2}$P$_{3/2}$}) pathway, and 0.17 and 0.23 eV for the I({\em $^{2}$P$_{1/2}$}) channel ({\em i.e.} weighted 4:3 averages of 0.24 and 0.33 eV for I$^{+}$ and CH$_2$Br$^{+}$, respectively). These values were used to model the curves in Fig.\ \ref{fig3}. The simulations in white assume the photofragments instantly reach their final velocities following photodissociation, and demonstrate that the two pathways are only separated by three to four camera pixels in their associated ion images. Improving the energy resolution of the experiment, for example by choosing different spectrometer conditions, would resolve these channels.

In the case of the I$^+$ fragment shown in the left panel of Fig.\ \ref{fig3}, it should be noted that there is good agreement with the model used at delay times shorter than 1.3\,ps. For longer delays, the kinetic energy release appears to be systematically lower than the simulation by around 0.1\,eV. The reason for this apparent discrepancy is that the delay-dependent Coulomb curve begins to intersect the low kinetic energy channel assigned to the production of I$^{+}$ through strong-field or UV multiple photon ionization. This overlap increases the observed intensity of the delay-dependent curve in the low kinetic energy region of the distribution. This needs to be borne in mind when comparing the fit and the experimental data. The Coulomb curve and the low kinetic energy channel are better separated for the CH$_2$Br$^{+}$ fragment shown in the right panel of Fig.\ \ref{fig3}, and so in this case the simulation agrees with the data at all time-delays.

The black curves in Fig.\ \ref{fig3} reconsider the assumption that the photofragments instantly reach their asymptotic velocities by modeling them with an exponential rise function and reevaluating their separation at each delay. This description approximates simple bond cleavage along a repulsive potential energy surface and should provide a more realistic photodissociation model without the computational expense of calculating the true molecular potential energy landscape. The time dependent velocity, $v(t)$, and bond length, $r_{\rm AB}(t)$ can be approximated by
\begin{equation}
v(t) = (v_{{\rm A},f}+v_{{\rm B},f}) \left( 1-{\rm e}^{-kt} \right)
\end{equation}
\begin{equation}
r_{{\rm AB}}(t) = \frac{v_{{\rm A},f}+v_{{\rm B},f}}{k} \left( kt+{\rm e}^{-kt}-1 \right) + r_{{\rm AB, eq}} \,. \end{equation}
In these equations, $v_{{\rm A},f}$ and $v_{{\rm B},f}$ are the final co-fragment velocities, $k$ is a dissociation rate constant, $t$ is the delay relative to $t_{0}$, and $r_{{\rm AB, eq}}$ is the equilibrium co-fragment separation before dissociation. A C-I dissociation lifetime ($\tau = 1/k$) of 50\,fs was previously determined by transient absorption spectroscopy \cite{Attar2014}. In the present work, the wave packet of the dissociating molecule is measured after its projection onto a Coulombic potential energy surface. The observed rate constant therefore represents a convolution of the dissociation and Coulomb explosion dynamics, as well as the {\em ca.}\ 170\,fs resolution of the experiment, and should indicate a slower process as a consequence. Using the model described above, a phenomenological lifetime of $240 \pm 60$\,fs was extracted from the delay-dependent I$^{+}$ and CH$_2$Br$^{+}$ channels and applied to simulate the black curves in Fig.\ \ref{fig3} (see Section IV). Additional experiments with shorter laser pulses would be required to ascertain which of the remaining factors, neutral dissociation dynamics or Coulomb explosion dynamics, contributes most to the parameter, $k$.

\begin{figure}
\includegraphics{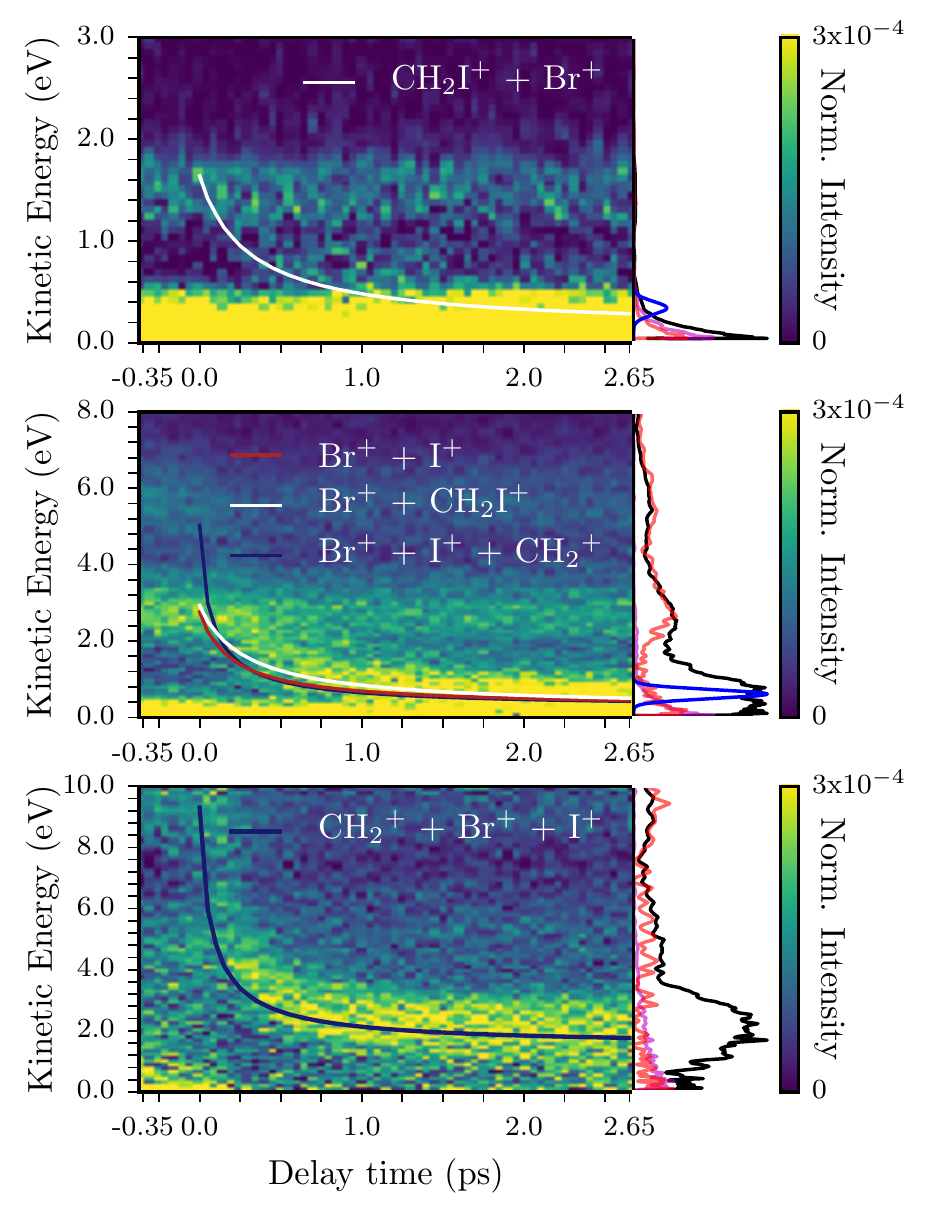}
\caption{CH$_2$I$^{+}$ (top), Br$^{+}$ (middle), and CH$_2$$^{+}$ (bottom) kinetic energy distributions, obtained from pBasex Abel inverted ion images, as a function of pump-probe delay. The 1.85\,ps data are projected to the right of each figure (black lines) and compared with normalized UV (violet lines) and IR background (red lines), and the sum of the asymptotic experimental data from Ref. \cite{Butler1987} and the electrostatic kinetic energy at 1.85\,ps (blue lines). Simulated curves for the expected kinetic energy release are labeled as X$^{+}$ + Y$^{+}$ + Z$^{+}$ for the plotted fragment X and co-fragments Y and Z.
\label{fig5}}
\end{figure}

Figure\ \ref{fig5} demonstrates the ability of timed Coulomb explosion imaging to investigate secondary reaction pathways by illustrating CH$_2$Br dissociation using the CH$_2$I$^{+}$, Br$^{+}$, and CH$_2$$^{+}$ kinetic energy distributions. The white curves simulate primary C-Br dissociation with the same constant fragment velocity model used in Fig.\ \ref{fig3} to demonstrate C-I cleavage. In this case, the CH$_2$I$^{+}$ distribution contains no obvious delay-dependent feature, while the Br$^{+}$ data exhibits two, one of which has comparable intensity to the CH$_2$Br$^{+}$ curve shown in Fig.\ \ref{fig3}. Considering that the CH$_2$I$^{+}$ ionization potential is 0.21\,eV lower than that of CH$_2$Br$^{+}$, and that a 4.8:1 C-I:C-Br fission branching ratio was previously observed at 266\,nm   \cite{Attar2014}, the absence of a delay-dependent CH$_2$I$^{+}$ channel confirms that primary C-Br cleavage is an unlikely dissociation pathway \cite{Andrews1984, Ma1993a}. The Br$^{+}$ curve therefore arises from a secondary or three-body process. Strong-field ionization following either of these processes could induce two- or three-body electrostatic repulsion and, as a consequence, a delay-dependent channel.

The expected three-body kinetic energy release of I$^{+}$, Br$^{+}$, and CH$_{2}$$^{+}$ is 18.2\,eV when using equilibrium internuclear distances of 345\,pm, 195\,pm, and 216\,pm for I-Br, C-Br, and C-I, respectively \cite{Liu2005, Kim2014, Bailleux2014b}. Summing the highest kinetic energy channels of Br$^{+}$ and CH$_{2}$$^{+}$ ($5.4 \pm 0.8$\,eV and $8.9 \pm 0.9$\,eV) with the diffuse I$^{+}$ feature centered at 3.3\,eV results in 17.6\,eV. However, three-body repulsion can be ruled out as a primary contributor to the delay-dependent Br$^{+}$ channel as the most intense region of the Br$^{+}$ curve appears to originate at 2.5\,eV rather than 5.4\,eV. The channel should therefore arise from two-body repulsion against I$^{+}$ or CH$_{2}$$^{+}$. The expected  Br$^{+}$ kinetic energies with these co-fragments in the negative delay regions are 2.57\,eV and 1.01\,eV, respectively, so the delay-dependent Br$^{+}$ channel is assigned to repulsion against I$^{+}$.

The production of I, Br, and CH$_{2}$ from CH$_{2}$BrI requires 5.66\,eV, which can be gained through the absorption of two 271.6\,nm photons, or from one UV photon and two 815\,nm photons. The asymptotic kinetic energies of the Br$^{+}$ and CH$_{2}$$^{+}$ curves suggest that two UV photons are absorbed. According to literature 11.3\% of the available energy internally excites CH$_{2}$ in this process, leaving 0.27\,eV and 1.51\,eV of translational energy for Br and CH$_{2}$, respectively, at 271.6\,nm \cite{Butler1987}. These neutral velocities lead to the red (two-body) and blue (three-body) simulated curves shown in Fig.\ 5 for Br$^{+}$ and CH$_{2}$$^{+}$, which match the delay-dependent channels and support the assignment detailed above.

\section{IV.\ Experimental Fitting}

For a two-body (AB) dissociation, the time-dependent fragment kinetic energies, $T_{{\rm A}}(t)$ and $T_{{\rm B}}(t)$, can also be described by the following equations:
\begin{eqnarray}
T_{{\rm A}}(t) &=& \frac{1}{2}{m_{\rm A}v_{\rm A}(t)^2} + \frac{k_{\rm e}q_{\rm A}q_{\rm B}}{r_{\rm AB}(t)} \left( \frac{m_{\rm B}}{m_{\rm A}+m_{\rm B}} \right) \nonumber \\   \\
T_{{\rm B}}(t) &=& \frac{1}{2}{m_{\rm B}v_{\rm B}(t)^2} + \frac{k_{\rm e}q_{\rm A}q_{\rm B}}{r_{\rm AB}(t)} \left( \frac{m_{\rm A}}{m_{\rm A}+m_{\rm B}} \right) \,. \nonumber
\end{eqnarray}
The total kinetic energy of fragments A and B, $T_{{\rm A}}(t-t_0)$ and $T_{{\rm B}}(t-t_0)$, determined by substitution of Eqs.\ (3) and (4) into (5), can therefore be described by four parameters: the final dissociation velocity, $v_{f} = v_{{\rm A},f} + v_{{\rm B},f}$, the rate constant, $k$, that characterizes the time-dependence of the velocities, the internuclear distance between co-fragments before ionization, $r_{{\rm AB, eq}}$, and the pump-probe delay, $t$\,-\,$t_{0}$. As was mentioned above, $k$ is a phenomenological rate constant parameter, which represents a convolution of the dissociation and Coulomb explosion dynamics, as well as the {\em ca.}\ 72\,fs uncertainty of the experiment.

\begin{figure*}
\includegraphics{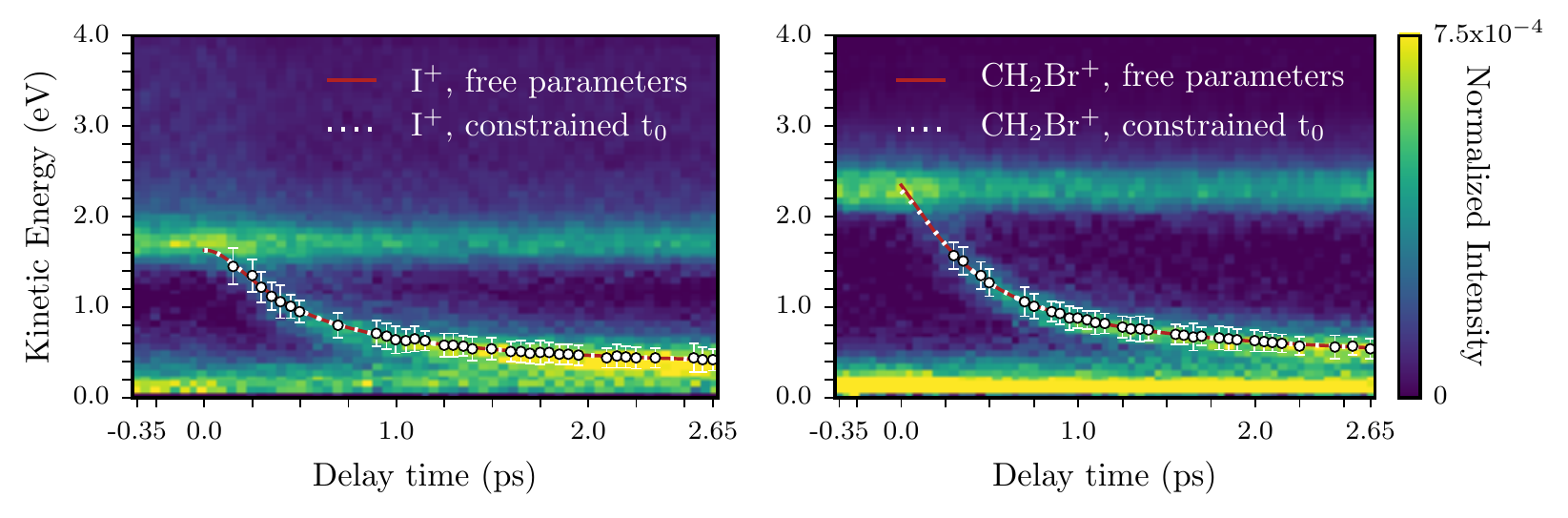}
\caption{I$^{+}$ (left) and CH$_{2}$Br$^{+}$ (right) kinetic energy distributions, obtained from pBasex Abel inverted ion images, as a function of pump-probe delay. The observed kinetic energies of the delay-dependent curves (white circles) are fit to Eq.\ (5) with and without (dashed white and solid red lines) a $t_{0}$ constraint of $\pm 100$\,fs to determine the parameters in Table 1.
\label{fig6}}
\end{figure*}

Figure\ \ref{fig6}  illustrates the result of fitting the expressions in Eq.\ (5) to the I$^{+}$ and CH$_{2}$Br$^{+}$ curves with and without a constraint applied to $t_{0}$. The associated parameters of each curve are collected in Table 1. Each variable can in principle be determined without restricting their range. In practice, however, both $t_{0}$ and the kinetic energy maxima at early pump-probe delays can be difficult to gauge. In the experiments reported here, $t_{0}$ was assigned using the cross-correlation intensity maxima of the I$^{+}$ and CH$_{2}$Br$^{+}$ fragments, and was restricted to $t_{0} \pm 100$\,fs. This roughly corresponds to a difference of two delay steps, and is comparable to the 72\,fs standard deviation of $t_{0}$. For I$^{+}$, the constrained and unconstrained curves are identical and are in excellent agreement with the assigned $t_{0}$. For CH$_{2}$Br$^{+}$, the lack of distinguishable kinetic energy maxima near $t_{0}$ leads to a small difference at the curve origin and, consequently, greater uncertainty in $t_{0}$, $r_{{\rm IBr, eq}}$ and $k$.

Table 1 also compares the accuracy of the reported parameters with CH$_{2}$BrI reference data. The established separation between charge centers at $t_{0}$ is within error of the expected I-Br equilibrium distance. The final neutral fragment velocities are slightly more difficult to compare. At 271.6\,nm, both ground (\textit{$^{2}$P$_{3/2}$}) and excited (\textit{$^{2}$P$_{1/2}$}) iodine spin-orbit states are created. As these states were not well resolved in this experiment, the reported kinetic energy maxima include contributions from both and represent the weighted average of the two states. I(\textit{$^{2}$P$_{3/2}$}) and I(\textit{$^{2}$P$_{1/2}$}) are present in a 4:3 ratio following photodissociation at 248\,nm and have respective final velocities centered around 0.30\,eV and 0.19\,eV \cite{Butler1987}. Their weighted average of 0.25\,eV matches the velocities reported here, and the same is true for the corresponding CH$_{2}$Br values. This ratio is expected to remain the same at 271.6\,nm, and so the internal energies reported by Lee and coworkers were used for the simulations presented above \cite{Parker1998, Senapati2002, Senapati2004}.

\begin{table}
\caption{\label{tables1}CH$_{2}$BrI photodissociation parameters at 271.6\,nm.}
\begin{ruledtabular}
\begin{tabular}{ccccccc}
  Fragment & Model & $t_{0}$\,(ps) & $k$\,(ps$^{-1}$) & $r_{{\rm IBr, eq}}$\,(pm) & $T_{{\rm I}}$\,(eV) & $T_{{\rm CH_{2}Br}}$\,(eV) \\
  \multirow{2}{*}{I$^{+}$} & Free parameters & 0.00 ${\pm}$ 0.06 & 4.1 ${\pm}$ 1.6 & 374 ${\pm}$ 26 & 0.265 ${\pm}$ 0.010 & 0.362 ${\pm}$ 0.013 \\
 & Constrained $t_{0}$  & 0.00 ${\pm}$ 0.06 & 4.1 ${\pm}$ 1.6 & 374 ${\pm}$ 26 & 0.265 ${\pm}$ 0.010 & 0.362 ${\pm}$ 0.013\\
\multirow{2}{*}{CH$_{2}$Br$^{+}$} & Free parameters & -0.20 ${\pm}$ 0.13 & 2.0 ${\pm}$ 1.0 & 310 ${\pm}$ 45 & 0.239 ${\pm}$ 0.007 & 0.327 ${\pm}$ 0.010 \\
 & Constrained $t_{0}$  & -0.10 ${\pm}$ 0.20 & 2.9 ${\pm}$ 2.4 & 350 ${\pm}$ 80 & 0.244 ${\pm}$ 0.017 & 0.333 ${\pm}$ 0.022\\
\hline
Reference data & & & 20.8\footnote{Reference \cite{Attar2014}} & 340-350\footnote{References \cite{Liu2005, Kim2014, Bailleux2014b}} & $0.25 \pm 0.14$\footnote{Reference \cite{Butler1987} (248\,nm)} & $0.34 \pm 0.15^{\rm c}$ \\
\end{tabular}
\end{ruledtabular}
\end{table}

\section{V. Conclusion}
Collectively, the delay-dependent fragment momenta illustrated in Figs.\ \ref{fig3} and \ref{fig5} demonstrate the concurrent dissociation dynamics of three processes. Single-photon dissociation of CH$_2$BrI at 271.6\,nm preferentially occurs through C-I cleavage, producing CH$_2$Br with I({\em $^{2}$P$_{3/2}$}) or I({\em $^{2}$P$_{1/2}$}); whereas absorption of a second UV photon breaks the C-Br bond of CH$_2$Br. The asymptotic kinetic energies of the primary fragments indicate that $65.6 \pm 1.6$\% of the available energy following C-I cleavage is deposited into internal CH$_2$Br states, matching the $67.9 \pm 1.5$\% reported at 248\,nm \cite{Butler1987}. Each delay-dependent channel was modeled using simple assumptions about the neutral dissociation and Coulomb explosion processes. The precision of timed Coulomb explosion imaging will improve as the stability and resolution of the acquisition hardware develop. In this work, a single PImMS pixel maps to a 0.04\,eV energy bin, and only 450 laser shots were taken per 50\,fs step. Improving these would further resolve the presented data.

With the above in mind, timed Coulomb explosion imaging could be directly applied to more complex photochemistry, including the dissociation of large molecules, multichromophoric compounds, or isomerization reactions. By using covariance or coincidence conditions, ion correlations could potentially be extracted \cite{Slater2015}, and information about the structural dynamics and chirality of parent molecules could be established.


\begin{acknowledgments}
\section{Acknowledgements}
The authors gratefully acknowledge the work of the scientific and technical team at FLASH, especially the FLASH laser group who made these experiments possible. The development of the CAMP instrument was funded by the Max Planck Society, and the installation of CAMP at FLASH was partially funded by BMBF grant No. 05K10KT2. The M.B., C.V., and S.R.M.\ groups are grateful for the support of the UK EPSRC (Programme Grant Nos.\ EP/G00224X/1 and EP/L005913/1), the EU (FP7 ITN `ICONIC', Project No.\ 238671), and the STFC (PNPAS award and  mini-IPS grant No.\ ST/J002895/1). A.Ru.\ and D.R.\ acknowledge support from the Chemical Sciences, Geosciences, and Biosciences Division, Office of Basic Energy Sciences, Office of Science, U.S.\ Department of Energy, Grant No.\ DE-FG02-86ER13491. D.R., E.S., R.B., C.B., and B.E.\ were also supported by the Helmholtz Gemeinschaft through the Helmholtz Young Investigator Program, and R.B.\ and S.Te.\ are additionally grateful for financial support from the German Research Council (DFG), CRC 755 (Project No.\ B03) and CRC 1073 (Project No.\ C02).  J.K. acknowledges support from the excellence cluster `The Hamburg Center for Ultrafast Imaging - Structure, Dynamics and Control of Matter at the Atomic Scale' of the Deutsche Forschungsgemeinschaft (CUI, DFG-EXC1074), the European Research Council under the European Union's Seventh Framework Programme (FP7/2007-2013) through the Consolidator Grant COMOTION (ERC-K\"upper-614507), and the Initiative and Networking Fund of the Helmholtz Association. S.B.\ thanks the Initiative and Networking Fund of the Helmholtz Association and the DFG (Grant No.\ SFB 755). The support of the EU to J.K., P.J., J.L., H.S., and D.R.\ {\em via} the MEDEA project within the Horizon 2020 research and innovation programme under the Marie Sk{\l}odowska-Curie grant agreement (No.\ 641789) is also gratefully acknowledged. A.S.M.\ and P.K.O.\ thank the German-Russian Interdisciplinary Science Center (G-RISC, C-2015a-6, C-2015b-6 and C-2016b-7) funded by the German Federal Foreign Office {\em via} the German Academic Exchange Service (DAAD). J.L., S.M.\ and P.J.\ thank the Swedish Research Council and the Swedish Foundation for Strategic Research. R.Ge. and T.R. acknowledge financial support from the French Agence Nationale de la Recherche (ANR) through the XSTASE project (ANR-14-CE32-0010).  B. C. M. is grateful for financial support from ANR SUMMIT. I. I. acknowledges financial support from Labex Plas@Par.
\end{acknowledgments}

%



\begin{thebibliography}{53}%
\makeatletter
\providecommand \@ifxundefined [1]{%
 \@ifx{#1\undefined}
}%
\providecommand \@ifnum [1]{%
 \ifnum #1\expandafter \@firstoftwo
 \else \expandafter \@secondoftwo
 \fi
}%
\providecommand \@ifx [1]{%
 \ifx #1\expandafter \@firstoftwo
 \else \expandafter \@secondoftwo
 \fi
}%
\providecommand \natexlab [1]{#1}%
\providecommand \enquote  [1]{``#1''}%
\providecommand \bibnamefont  [1]{#1}%
\providecommand \bibfnamefont [1]{#1}%
\providecommand \citenamefont [1]{#1}%
\providecommand \href@noop [0]{\@secondoftwo}%
\providecommand \href [0]{\begingroup \@sanitize@url \@href}%
\providecommand \@href[1]{\@@startlink{#1}\@@href}%
\providecommand \@@href[1]{\endgroup#1\@@endlink}%
\providecommand \@sanitize@url [0]{\catcode `\\12\catcode `\$12\catcode
  `\&12\catcode `\#12\catcode `\^12\catcode `\_12\catcode `\%12\relax}%
\providecommand \@@startlink[1]{}%
\providecommand \@@endlink[0]{}%
\providecommand \url  [0]{\begingroup\@sanitize@url \@url }%
\providecommand \@url [1]{\endgroup\@href {#1}{\urlprefix }}%
\providecommand \urlprefix  [0]{URL }%
\providecommand \Eprint [0]{\href }%
\providecommand \doibase [0]{http://dx.doi.org/}%
\providecommand \selectlanguage [0]{\@gobble}%
\providecommand \bibinfo  [0]{\@secondoftwo}%
\providecommand \bibfield  [0]{\@secondoftwo}%
\providecommand \translation [1]{[#1]}%
\providecommand \BibitemOpen [0]{}%
\providecommand \bibitemStop [0]{}%
\providecommand \bibitemNoStop [0]{.\EOS\space}%
\providecommand \EOS [0]{\spacefactor3000\relax}%
\providecommand \BibitemShut  [1]{\csname bibitem#1\endcsname}%
\let\auto@bib@innerbib\@empty
\bibitem [{\citenamefont {Zewail}(1988)}]{Zewail1988}%
  \BibitemOpen
  \bibfield  {author} {\bibinfo {author} {\bibfnamefont {A.~H.}\ \bibnamefont
  {Zewail}},\ }\href {\doibase 10.1002/job.322} {\bibfield  {journal} {\bibinfo
   {journal} {Science}\ }\textbf {\bibinfo {volume} {242}},\ \bibinfo {pages}
  {1645} (\bibinfo {year} {1988})}\BibitemShut {NoStop}%
\bibitem [{\citenamefont {Dur{\'{a}}}\ \emph {et~al.}(2008)\citenamefont
  {Dur{\'{a}}}, \citenamefont {{de Nalda}}, \citenamefont {{\'{A}}lvarez},
  \citenamefont {Izquierdo}, \citenamefont {Amaral},\ and\ \citenamefont
  {Ba{\~{n}}ares}}]{Dura2008}%
  \BibitemOpen
  \bibfield  {author} {\bibinfo {author} {\bibfnamefont {J.}~\bibnamefont
  {Dur{\'{a}}}}, \bibinfo {author} {\bibfnamefont {R.}~\bibnamefont {{de
  Nalda}}}, \bibinfo {author} {\bibfnamefont {J.}~\bibnamefont
  {{\'{A}}lvarez}}, \bibinfo {author} {\bibfnamefont {J.~G.}\ \bibnamefont
  {Izquierdo}}, \bibinfo {author} {\bibfnamefont {G.~A.}\ \bibnamefont
  {Amaral}}, \ and\ \bibinfo {author} {\bibfnamefont {L.}~\bibnamefont
  {Ba{\~{n}}ares}},\ }\href {\doibase 10.1002/cphc.200800018} {\bibfield
  {journal} {\bibinfo  {journal} {ChemPhysChem}\ }\textbf {\bibinfo {volume}
  {9}},\ \bibinfo {pages} {1245} (\bibinfo {year} {2008})}\BibitemShut
  {NoStop}%
\bibitem [{\citenamefont {de~Nalda}\ \emph {et~al.}(2008)\citenamefont
  {de~Nalda}, \citenamefont {Dur{\'{a}}}, \citenamefont {Garc{\'{i}}a-Vela},
  \citenamefont {Izquierdo}, \citenamefont {Gonz{\'{a}}lez-V{\'{a}}zquez},\
  and\ \citenamefont {Ba{\~{n}}ares}}]{DeNalda2008}%
  \BibitemOpen
  \bibfield  {author} {\bibinfo {author} {\bibfnamefont {R.}~\bibnamefont
  {de~Nalda}}, \bibinfo {author} {\bibfnamefont {J.}~\bibnamefont
  {Dur{\'{a}}}}, \bibinfo {author} {\bibfnamefont {A.}~\bibnamefont
  {Garc{\'{i}}a-Vela}}, \bibinfo {author} {\bibfnamefont {J.~G.}\ \bibnamefont
  {Izquierdo}}, \bibinfo {author} {\bibfnamefont {J.}~\bibnamefont
  {Gonz{\'{a}}lez-V{\'{a}}zquez}}, \ and\ \bibinfo {author} {\bibfnamefont
  {L.}~\bibnamefont {Ba{\~{n}}ares}},\ }\href {\doibase 10.1063/1.2943198}
  {\bibfield  {journal} {\bibinfo  {journal} {J. Chem. Phys.}\ }\textbf
  {\bibinfo {volume} {128}},\ \bibinfo {pages} {244309} (\bibinfo {year}
  {2008})}\BibitemShut {NoStop}%
\bibitem [{\citenamefont {Corrales}\ \emph {et~al.}(2014)\citenamefont
  {Corrales}, \citenamefont {Gonz{\'{a}}lez-V{\'{a}}zquez}, \citenamefont
  {Balerdi}, \citenamefont {Sol{\'{a}}}, \citenamefont {de~Nalda},\ and\
  \citenamefont {Ba{\~{n}}ares}}]{Corrales2014}%
  \BibitemOpen
  \bibfield  {author} {\bibinfo {author} {\bibfnamefont {M.~E.}\ \bibnamefont
  {Corrales}}, \bibinfo {author} {\bibfnamefont {J.}~\bibnamefont
  {Gonz{\'{a}}lez-V{\'{a}}zquez}}, \bibinfo {author} {\bibfnamefont
  {G.}~\bibnamefont {Balerdi}}, \bibinfo {author} {\bibfnamefont {I.~R.}\
  \bibnamefont {Sol{\'{a}}}}, \bibinfo {author} {\bibfnamefont
  {R.}~\bibnamefont {de~Nalda}}, \ and\ \bibinfo {author} {\bibfnamefont
  {L.}~\bibnamefont {Ba{\~{n}}ares}},\ }\href {\doibase 10.1038/nchem.2006}
  {\bibfield  {journal} {\bibinfo  {journal} {Nat. Chem.}\ }\textbf {\bibinfo
  {volume} {6}},\ \bibinfo {pages} {785} (\bibinfo {year} {2014})}\BibitemShut
  {NoStop}%
\bibitem [{\citenamefont {Ibrahim}\ \emph {et~al.}(2014)\citenamefont
  {Ibrahim}, \citenamefont {Wales}, \citenamefont {Beaulieu}, \citenamefont
  {Schmidt}, \citenamefont {Thir{\'{e}}}, \citenamefont {Fowe}, \citenamefont
  {Bisson}, \citenamefont {Hebeisen}, \citenamefont {Wanie}, \citenamefont
  {Gigu{\'{e}}re}, \citenamefont {Kieffer}, \citenamefont {Spanner},
  \citenamefont {Bandrauk}, \citenamefont {Sanderson}, \citenamefont
  {Schuurman},\ and\ \citenamefont {L{\'{e}}gar{\'{e}}}}]{Ibrahim2014}%
  \BibitemOpen
  \bibfield  {author} {\bibinfo {author} {\bibfnamefont {H.}~\bibnamefont
  {Ibrahim}}, \bibinfo {author} {\bibfnamefont {B.}~\bibnamefont {Wales}},
  \bibinfo {author} {\bibfnamefont {S.}~\bibnamefont {Beaulieu}}, \bibinfo
  {author} {\bibfnamefont {B.~E.}\ \bibnamefont {Schmidt}}, \bibinfo {author}
  {\bibfnamefont {N.}~\bibnamefont {Thir{\'{e}}}}, \bibinfo {author}
  {\bibfnamefont {E.~P.}\ \bibnamefont {Fowe}}, \bibinfo {author}
  {\bibfnamefont {{\'{E}}.}~\bibnamefont {Bisson}}, \bibinfo {author}
  {\bibfnamefont {C.~T.}\ \bibnamefont {Hebeisen}}, \bibinfo {author}
  {\bibfnamefont {V.}~\bibnamefont {Wanie}}, \bibinfo {author} {\bibfnamefont
  {M.}~\bibnamefont {Gigu{\'{e}}re}}, \bibinfo {author} {\bibfnamefont {J.-C.}\
  \bibnamefont {Kieffer}}, \bibinfo {author} {\bibfnamefont {M.}~\bibnamefont
  {Spanner}}, \bibinfo {author} {\bibfnamefont {A.~D.}\ \bibnamefont
  {Bandrauk}}, \bibinfo {author} {\bibfnamefont {J.}~\bibnamefont {Sanderson}},
  \bibinfo {author} {\bibfnamefont {M.~S.}\ \bibnamefont {Schuurman}}, \ and\
  \bibinfo {author} {\bibfnamefont {F.}~\bibnamefont {L{\'{e}}gar{\'{e}}}},\
  }\href {\doibase 10.1038/ncomms5422} {\bibfield  {journal} {\bibinfo
  {journal} {Nat. Commun.}\ }\textbf {\bibinfo {volume} {5}},\ \bibinfo {pages}
  {4422} (\bibinfo {year} {2014})}\BibitemShut {NoStop}%
\bibitem [{\citenamefont {Vager}\ \emph {et~al.}(1989)\citenamefont {Vager},
  \citenamefont {Naaman},\ and\ \citenamefont {Kanter}}]{Vager1989a}%
  \BibitemOpen
  \bibfield  {author} {\bibinfo {author} {\bibfnamefont {Z.}~\bibnamefont
  {Vager}}, \bibinfo {author} {\bibfnamefont {R.}~\bibnamefont {Naaman}}, \
  and\ \bibinfo {author} {\bibfnamefont {E.~P.}\ \bibnamefont {Kanter}},\
  }\href@noop {} {\bibfield  {journal} {\bibinfo  {journal} {Science}\ }\textbf
  {\bibinfo {volume} {244}},\ \bibinfo {pages} {426} (\bibinfo {year}
  {1989})}\BibitemShut {NoStop}%
\bibitem [{\citenamefont {Cornaggia}\ \emph {et~al.}(1992)\citenamefont
  {Cornaggia}, \citenamefont {Normand},\ and\ \citenamefont
  {Morellec}}]{Cornaggia1992}%
  \BibitemOpen
  \bibfield  {author} {\bibinfo {author} {\bibfnamefont {C.}~\bibnamefont
  {Cornaggia}}, \bibinfo {author} {\bibfnamefont {D.}~\bibnamefont {Normand}},
  \ and\ \bibinfo {author} {\bibfnamefont {J.}~\bibnamefont {Morellec}},\
  }\href {\doibase 10.1088/0953-4075/25/17/003} {\bibfield  {journal} {\bibinfo
   {journal} {J. Phys. B}\ }\textbf {\bibinfo {volume} {25}},\ \bibinfo {pages}
  {L415} (\bibinfo {year} {1992})}\BibitemShut {NoStop}%
\bibitem [{\citenamefont {Hering}\ and\ \citenamefont
  {Cornaggia}(1999)}]{Hering1999}%
  \BibitemOpen
  \bibfield  {author} {\bibinfo {author} {\bibfnamefont {P.}~\bibnamefont
  {Hering}}\ and\ \bibinfo {author} {\bibfnamefont {C.}~\bibnamefont
  {Cornaggia}},\ }\href {\doibase 10.1103/PhysRevA.59.2836} {\bibfield
  {journal} {\bibinfo  {journal} {Phys. Rev. A}\ }\textbf {\bibinfo {volume}
  {59}},\ \bibinfo {pages} {2836} (\bibinfo {year} {1999})}\BibitemShut
  {NoStop}%
\bibitem [{\citenamefont {Sanderson}\ \emph {et~al.}(1999)\citenamefont
  {Sanderson}, \citenamefont {El-Zein}, \citenamefont {Bryan}, \citenamefont
  {Newell}, \citenamefont {Langley},\ and\ \citenamefont
  {Taday}}]{Sanderson1999}%
  \BibitemOpen
  \bibfield  {author} {\bibinfo {author} {\bibfnamefont {J.~H.}\ \bibnamefont
  {Sanderson}}, \bibinfo {author} {\bibfnamefont {A.}~\bibnamefont {El-Zein}},
  \bibinfo {author} {\bibfnamefont {W.~A.}\ \bibnamefont {Bryan}}, \bibinfo
  {author} {\bibfnamefont {W.~R.}\ \bibnamefont {Newell}}, \bibinfo {author}
  {\bibfnamefont {A.~J.}\ \bibnamefont {Langley}}, \ and\ \bibinfo {author}
  {\bibfnamefont {P.~F.}\ \bibnamefont {Taday}},\ }\href {\doibase
  10.1103/PhysRevA.59.R2567} {\bibfield  {journal} {\bibinfo  {journal} {Phys.
  Rev. A}\ }\textbf {\bibinfo {volume} {59}},\ \bibinfo {pages} {R2567}
  (\bibinfo {year} {1999})}\BibitemShut {NoStop}%
\bibitem [{\citenamefont {L{\'{e}}gar{\'{e}}}\ \emph
  {et~al.}(2005)\citenamefont {L{\'{e}}gar{\'{e}}}, \citenamefont {Lee},
  \citenamefont {Litvinyuk}, \citenamefont {Dooley}, \citenamefont
  {Wesolowski}, \citenamefont {Bunker}, \citenamefont {Dombi}, \citenamefont
  {Krausz}, \citenamefont {Bandrauk}, \citenamefont {Villeneuve},\ and\
  \citenamefont {Corkum}}]{Legare2005}%
  \BibitemOpen
  \bibfield  {author} {\bibinfo {author} {\bibfnamefont {F.}~\bibnamefont
  {L{\'{e}}gar{\'{e}}}}, \bibinfo {author} {\bibfnamefont {K.~F.}\ \bibnamefont
  {Lee}}, \bibinfo {author} {\bibfnamefont {I.~V.}\ \bibnamefont {Litvinyuk}},
  \bibinfo {author} {\bibfnamefont {P.~W.}\ \bibnamefont {Dooley}}, \bibinfo
  {author} {\bibfnamefont {S.~S.}\ \bibnamefont {Wesolowski}}, \bibinfo
  {author} {\bibfnamefont {P.~R.}\ \bibnamefont {Bunker}}, \bibinfo {author}
  {\bibfnamefont {P.}~\bibnamefont {Dombi}}, \bibinfo {author} {\bibfnamefont
  {F.}~\bibnamefont {Krausz}}, \bibinfo {author} {\bibfnamefont {A.~D.}\
  \bibnamefont {Bandrauk}}, \bibinfo {author} {\bibfnamefont {D.~M.}\
  \bibnamefont {Villeneuve}}, \ and\ \bibinfo {author} {\bibfnamefont {P.~B.}\
  \bibnamefont {Corkum}},\ }\href {\doibase 10.1103/PhysRevA.71.013415}
  {\bibfield  {journal} {\bibinfo  {journal} {Phys. Rev. A}\ }\textbf {\bibinfo
  {volume} {71}},\ \bibinfo {pages} {013415} (\bibinfo {year}
  {2005})}\BibitemShut {NoStop}%
\bibitem [{\citenamefont {Gagnon}\ \emph {et~al.}(2008)\citenamefont {Gagnon},
  \citenamefont {Lee}, \citenamefont {Rayner}, \citenamefont {Corkum},\ and\
  \citenamefont {Bhardwaj}}]{Gagnon2008}%
  \BibitemOpen
  \bibfield  {author} {\bibinfo {author} {\bibfnamefont {J.}~\bibnamefont
  {Gagnon}}, \bibinfo {author} {\bibfnamefont {K.~F.}\ \bibnamefont {Lee}},
  \bibinfo {author} {\bibfnamefont {D.~M.}\ \bibnamefont {Rayner}}, \bibinfo
  {author} {\bibfnamefont {P.~B.}\ \bibnamefont {Corkum}}, \ and\ \bibinfo
  {author} {\bibfnamefont {V.~R.}\ \bibnamefont {Bhardwaj}},\ }\href {\doibase
  10.1088/0953-4075/41/21/215104} {\bibfield  {journal} {\bibinfo  {journal}
  {J. Phys. B}\ }\textbf {\bibinfo {volume} {41}},\ \bibinfo {pages} {215104}
  (\bibinfo {year} {2008})}\BibitemShut {NoStop}%
\bibitem [{\citenamefont {Corrales}\ \emph {et~al.}(2012)\citenamefont
  {Corrales}, \citenamefont {Gitzinger}, \citenamefont
  {Gonz{\'{a}}lez-V{\'{a}}zquez}, \citenamefont {Loriot}, \citenamefont {{de
  Nalda}},\ and\ \citenamefont {Ba{\~{n}}ares}}]{Corrales2012}%
  \BibitemOpen
  \bibfield  {author} {\bibinfo {author} {\bibfnamefont {M.~E.}\ \bibnamefont
  {Corrales}}, \bibinfo {author} {\bibfnamefont {G.}~\bibnamefont {Gitzinger}},
  \bibinfo {author} {\bibfnamefont {J.}~\bibnamefont
  {Gonz{\'{a}}lez-V{\'{a}}zquez}}, \bibinfo {author} {\bibfnamefont
  {V.}~\bibnamefont {Loriot}}, \bibinfo {author} {\bibfnamefont
  {R.}~\bibnamefont {{de Nalda}}}, \ and\ \bibinfo {author} {\bibfnamefont
  {L.}~\bibnamefont {Ba{\~{n}}ares}},\ }\href {\doibase 10.1021/jp207367a}
  {\bibfield  {journal} {\bibinfo  {journal} {J. Phys. Chem. A}\ }\textbf
  {\bibinfo {volume} {116}},\ \bibinfo {pages} {2669} (\bibinfo {year}
  {2012})}\BibitemShut {NoStop}%
\bibitem [{\citenamefont {Pitzer}\ \emph {et~al.}(2013)\citenamefont {Pitzer},
  \citenamefont {Kunitski}, \citenamefont {Johnson}, \citenamefont {Jahnke},
  \citenamefont {Sann}, \citenamefont {Sturm}, \citenamefont {Schmidt},
  \citenamefont {Schmidt-B{\"{o}}cking}, \citenamefont {D{\"{o}}rner},
  \citenamefont {Stohner}, \citenamefont {Kiedrowski}, \citenamefont
  {Reggelin}, \citenamefont {Marquardt}, \citenamefont {Schie$\beta$er},
  \citenamefont {Berger},\ and\ \citenamefont {Sch{\"{o}}ffler}}]{Pitzer2013}%
  \BibitemOpen
  \bibfield  {author} {\bibinfo {author} {\bibfnamefont {M.}~\bibnamefont
  {Pitzer}}, \bibinfo {author} {\bibfnamefont {M.}~\bibnamefont {Kunitski}},
  \bibinfo {author} {\bibfnamefont {A.~S.}\ \bibnamefont {Johnson}}, \bibinfo
  {author} {\bibfnamefont {T.}~\bibnamefont {Jahnke}}, \bibinfo {author}
  {\bibfnamefont {H.}~\bibnamefont {Sann}}, \bibinfo {author} {\bibfnamefont
  {F.}~\bibnamefont {Sturm}}, \bibinfo {author} {\bibfnamefont {L.~P.~H.}\
  \bibnamefont {Schmidt}}, \bibinfo {author} {\bibfnamefont {H.}~\bibnamefont
  {Schmidt-B{\"{o}}cking}}, \bibinfo {author} {\bibfnamefont {R.}~\bibnamefont
  {D{\"{o}}rner}}, \bibinfo {author} {\bibfnamefont {J.}~\bibnamefont
  {Stohner}}, \bibinfo {author} {\bibfnamefont {J.}~\bibnamefont {Kiedrowski}},
  \bibinfo {author} {\bibfnamefont {M.}~\bibnamefont {Reggelin}}, \bibinfo
  {author} {\bibfnamefont {S.}~\bibnamefont {Marquardt}}, \bibinfo {author}
  {\bibfnamefont {A.}~\bibnamefont {Schie$\beta$er}}, \bibinfo {author}
  {\bibfnamefont {R.}~\bibnamefont {Berger}}, \ and\ \bibinfo {author}
  {\bibfnamefont {M.~S.}\ \bibnamefont {Sch{\"{o}}ffler}},\ }\href@noop {}
  {\bibfield  {journal} {\bibinfo  {journal} {Science}\ }\textbf {\bibinfo
  {volume} {341}},\ \bibinfo {pages} {1096} (\bibinfo {year}
  {2013})}\BibitemShut {NoStop}%
\bibitem [{\citenamefont {Slater}\ \emph {et~al.}(2014)\citenamefont {Slater},
  \citenamefont {Blake}, \citenamefont {Brouard}, \citenamefont {Lauer},
  \citenamefont {Vallance}, \citenamefont {John}, \citenamefont {Turchetta},
  \citenamefont {Nomerotski}, \citenamefont {Christensen}, \citenamefont
  {Nielsen}, \citenamefont {Johansson},\ and\ \citenamefont
  {Stapelfeldt}}]{Slater2014}%
  \BibitemOpen
  \bibfield  {author} {\bibinfo {author} {\bibfnamefont {C.~S.}\ \bibnamefont
  {Slater}}, \bibinfo {author} {\bibfnamefont {S.}~\bibnamefont {Blake}},
  \bibinfo {author} {\bibfnamefont {M.}~\bibnamefont {Brouard}}, \bibinfo
  {author} {\bibfnamefont {A.}~\bibnamefont {Lauer}}, \bibinfo {author}
  {\bibfnamefont {C.}~\bibnamefont {Vallance}}, \bibinfo {author}
  {\bibfnamefont {J.~J.}\ \bibnamefont {John}}, \bibinfo {author}
  {\bibfnamefont {R.}~\bibnamefont {Turchetta}}, \bibinfo {author}
  {\bibfnamefont {A.}~\bibnamefont {Nomerotski}}, \bibinfo {author}
  {\bibfnamefont {L.}~\bibnamefont {Christensen}}, \bibinfo {author}
  {\bibfnamefont {J.~H.}\ \bibnamefont {Nielsen}}, \bibinfo {author}
  {\bibfnamefont {M.~P.}\ \bibnamefont {Johansson}}, \ and\ \bibinfo {author}
  {\bibfnamefont {H.}~\bibnamefont {Stapelfeldt}},\ }\href {\doibase
  10.1103/PhysRevA.89.011401} {\bibfield  {journal} {\bibinfo  {journal} {Phys.
  Rev. A}\ }\textbf {\bibinfo {volume} {89}},\ \bibinfo {pages} {011401}
  (\bibinfo {year} {2014})}\BibitemShut {NoStop}%
\bibitem [{\citenamefont {Christensen}\ \emph {et~al.}(2015)\citenamefont
  {Christensen}, \citenamefont {Nielsen}, \citenamefont {Slater}, \citenamefont
  {Lauer}, \citenamefont {Brouard},\ and\ \citenamefont
  {Stapelfeldt}}]{Christensen2015}%
  \BibitemOpen
  \bibfield  {author} {\bibinfo {author} {\bibfnamefont {L.}~\bibnamefont
  {Christensen}}, \bibinfo {author} {\bibfnamefont {J.~H.}\ \bibnamefont
  {Nielsen}}, \bibinfo {author} {\bibfnamefont {C.~S.}\ \bibnamefont {Slater}},
  \bibinfo {author} {\bibfnamefont {A.}~\bibnamefont {Lauer}}, \bibinfo
  {author} {\bibfnamefont {M.}~\bibnamefont {Brouard}}, \ and\ \bibinfo
  {author} {\bibfnamefont {H.}~\bibnamefont {Stapelfeldt}},\ }\href {\doibase
  10.1103/PhysRevA.92.033411} {\bibfield  {journal} {\bibinfo  {journal} {Phys.
  Rev. A}\ }\textbf {\bibinfo {volume} {92}},\ \bibinfo {pages} {033411}
  (\bibinfo {year} {2015})}\BibitemShut {NoStop}%
\bibitem [{\citenamefont {Christiansen}\ \emph {et~al.}(2016)\citenamefont
  {Christiansen}, \citenamefont {Nielsen}, \citenamefont {Christensen},
  \citenamefont {Shepperson}, \citenamefont {Pentlehner},\ and\ \citenamefont
  {Stapelfeldt}}]{Christiansen2016}%
  \BibitemOpen
  \bibfield  {author} {\bibinfo {author} {\bibfnamefont {L.}~\bibnamefont
  {Christiansen}}, \bibinfo {author} {\bibfnamefont {J.~H.}\ \bibnamefont
  {Nielsen}}, \bibinfo {author} {\bibfnamefont {L.}~\bibnamefont
  {Christensen}}, \bibinfo {author} {\bibfnamefont {B.}~\bibnamefont
  {Shepperson}}, \bibinfo {author} {\bibfnamefont {D.}~\bibnamefont
  {Pentlehner}}, \ and\ \bibinfo {author} {\bibfnamefont {H.}~\bibnamefont
  {Stapelfeldt}},\ }\href {\doibase 10.1103/PhysRevA.93.023411} {\bibfield
  {journal} {\bibinfo  {journal} {Phys. Rev. A}\ }\textbf {\bibinfo {volume}
  {93}},\ \bibinfo {pages} {023411} (\bibinfo {year} {2016})}\BibitemShut
  {NoStop}%
\bibitem [{\citenamefont {Frasinski}\ \emph {et~al.}(1989)\citenamefont
  {Frasinski}, \citenamefont {Codling},\ and\ \citenamefont
  {Hatherly}}]{Frasinski1989}%
  \BibitemOpen
  \bibfield  {author} {\bibinfo {author} {\bibfnamefont {L.~J.}\ \bibnamefont
  {Frasinski}}, \bibinfo {author} {\bibfnamefont {K.}~\bibnamefont {Codling}},
  \ and\ \bibinfo {author} {\bibfnamefont {P.~A.}\ \bibnamefont {Hatherly}},\
  }\href {\doibase 10.1126/science.246.4933.1029} {\bibfield  {journal}
  {\bibinfo  {journal} {Science}\ }\textbf {\bibinfo {volume} {246}},\ \bibinfo
  {pages} {1029} (\bibinfo {year} {1989})}\BibitemShut {NoStop}%
\bibitem [{\citenamefont {Frasinski}(2016)}]{Frasinski2016}%
  \BibitemOpen
  \bibfield  {author} {\bibinfo {author} {\bibfnamefont {L.~J.}\ \bibnamefont
  {Frasinski}},\ }\href {\doibase 10.1088/0953-4075/49/15/152004} {\bibfield
  {journal} {\bibinfo  {journal} {J. Phys. B}\ }\textbf {\bibinfo {volume}
  {49}},\ \bibinfo {pages} {152004} (\bibinfo {year} {2016})}\BibitemShut
  {NoStop}%
\bibitem [{\citenamefont {Kitamura}\ \emph {et~al.}(2001)\citenamefont
  {Kitamura}, \citenamefont {Nishide}, \citenamefont {Shiromaru}, \citenamefont
  {Achiba},\ and\ \citenamefont {Kobayashi}}]{Kitamura2001}%
  \BibitemOpen
  \bibfield  {author} {\bibinfo {author} {\bibfnamefont {T.}~\bibnamefont
  {Kitamura}}, \bibinfo {author} {\bibfnamefont {T.}~\bibnamefont {Nishide}},
  \bibinfo {author} {\bibfnamefont {H.}~\bibnamefont {Shiromaru}}, \bibinfo
  {author} {\bibfnamefont {Y.}~\bibnamefont {Achiba}}, \ and\ \bibinfo {author}
  {\bibfnamefont {N.}~\bibnamefont {Kobayashi}},\ }\href {\doibase
  10.1063/1.1383793} {\bibfield  {journal} {\bibinfo  {journal} {J. Chem.
  Phys.}\ }\textbf {\bibinfo {volume} {115}},\ \bibinfo {pages} {5} (\bibinfo
  {year} {2001})}\BibitemShut {NoStop}%
\bibitem [{\citenamefont {Pickering}\ \emph {et~al.}(2016)\citenamefont
  {Pickering}, \citenamefont {Amini}, \citenamefont {Brouard}, \citenamefont
  {Burt}, \citenamefont {Bush}, \citenamefont {Christensen}, \citenamefont
  {Lauer}, \citenamefont {Nielsen}, \citenamefont {Slater},\ and\ \citenamefont
  {Stapelfeldt}}]{Pickering2016}%
  \BibitemOpen
  \bibfield  {author} {\bibinfo {author} {\bibfnamefont {J.~D.}\ \bibnamefont
  {Pickering}}, \bibinfo {author} {\bibfnamefont {K.}~\bibnamefont {Amini}},
  \bibinfo {author} {\bibfnamefont {M.}~\bibnamefont {Brouard}}, \bibinfo
  {author} {\bibfnamefont {M.}~\bibnamefont {Burt}}, \bibinfo {author}
  {\bibfnamefont {I.~J.}\ \bibnamefont {Bush}}, \bibinfo {author}
  {\bibfnamefont {L.}~\bibnamefont {Christensen}}, \bibinfo {author}
  {\bibfnamefont {A.}~\bibnamefont {Lauer}}, \bibinfo {author} {\bibfnamefont
  {J.~H.}\ \bibnamefont {Nielsen}}, \bibinfo {author} {\bibfnamefont {C.~S.}\
  \bibnamefont {Slater}}, \ and\ \bibinfo {author} {\bibfnamefont
  {H.}~\bibnamefont {Stapelfeldt}},\ }\href {\doibase 10.1063/1.4947551}
  {\bibfield  {journal} {\bibinfo  {journal} {J. Chem. Phys.}\ }\textbf
  {\bibinfo {volume} {144}},\ \bibinfo {pages} {161105} (\bibinfo {year}
  {2016})}\BibitemShut {NoStop}%
\bibitem [{\citenamefont {Stapelfeldt}\ \emph {et~al.}(1995)\citenamefont
  {Stapelfeldt}, \citenamefont {Constant},\ and\ \citenamefont
  {Corkum}}]{Stapelfeldt1995}%
  \BibitemOpen
  \bibfield  {author} {\bibinfo {author} {\bibfnamefont {H.}~\bibnamefont
  {Stapelfeldt}}, \bibinfo {author} {\bibfnamefont {E.}~\bibnamefont
  {Constant}}, \ and\ \bibinfo {author} {\bibfnamefont {P.~B.}\ \bibnamefont
  {Corkum}},\ }\href {\doibase 10.1103/PhysRevLett.74.3780} {\bibfield
  {journal} {\bibinfo  {journal} {Phys. Rev. Lett.}\ }\textbf {\bibinfo
  {volume} {74}},\ \bibinfo {pages} {3780} (\bibinfo {year}
  {1995})}\BibitemShut {NoStop}%
\bibitem [{\citenamefont {Ergler}\ \emph {et~al.}(2005)\citenamefont {Ergler},
  \citenamefont {Rudenko}, \citenamefont {Feuerstein}, \citenamefont {Zrost},
  \citenamefont {Schr{\"{o}}ter}, \citenamefont {Moshammer},\ and\
  \citenamefont {Ullrich}}]{Ergler2005a}%
  \BibitemOpen
  \bibfield  {author} {\bibinfo {author} {\bibfnamefont {T.}~\bibnamefont
  {Ergler}}, \bibinfo {author} {\bibfnamefont {A.}~\bibnamefont {Rudenko}},
  \bibinfo {author} {\bibfnamefont {B.}~\bibnamefont {Feuerstein}}, \bibinfo
  {author} {\bibfnamefont {K.}~\bibnamefont {Zrost}}, \bibinfo {author}
  {\bibfnamefont {C.~D.}\ \bibnamefont {Schr{\"{o}}ter}}, \bibinfo {author}
  {\bibfnamefont {R.}~\bibnamefont {Moshammer}}, \ and\ \bibinfo {author}
  {\bibfnamefont {J.}~\bibnamefont {Ullrich}},\ }\href {\doibase
  10.1103/PhysRevLett.95.093001} {\bibfield  {journal} {\bibinfo  {journal}
  {Phys. Rev. Lett.}\ }\textbf {\bibinfo {volume} {95}},\ \bibinfo {pages}
  {093001} (\bibinfo {year} {2005})}\BibitemShut {NoStop}%
\bibitem [{\citenamefont {L{\'{e}}gar{\'{e}}}\ \emph
  {et~al.}(2006)\citenamefont {L{\'{e}}gar{\'{e}}}, \citenamefont {Lee},
  \citenamefont {Bandrauk}, \citenamefont {Villeneuve},\ and\ \citenamefont
  {Corkum}}]{Legare2006b}%
  \BibitemOpen
  \bibfield  {author} {\bibinfo {author} {\bibfnamefont {F.}~\bibnamefont
  {L{\'{e}}gar{\'{e}}}}, \bibinfo {author} {\bibfnamefont {K.~F.}\ \bibnamefont
  {Lee}}, \bibinfo {author} {\bibfnamefont {A.~D.}\ \bibnamefont {Bandrauk}},
  \bibinfo {author} {\bibfnamefont {D.~M.}\ \bibnamefont {Villeneuve}}, \ and\
  \bibinfo {author} {\bibfnamefont {P.~B.}\ \bibnamefont {Corkum}},\ }\href
  {\doibase 10.1088/0953-4075/39/13/S23} {\bibfield  {journal} {\bibinfo
  {journal} {J. Phys. B}\ }\textbf {\bibinfo {volume} {39}},\ \bibinfo {pages}
  {S503} (\bibinfo {year} {2006})}\BibitemShut {NoStop}%
\bibitem [{\citenamefont {Christensen}\ \emph {et~al.}(2014)\citenamefont
  {Christensen}, \citenamefont {Nielsen}, \citenamefont {Brandt}, \citenamefont
  {Madsen}, \citenamefont {Madsen}, \citenamefont {Slater}, \citenamefont
  {Lauer}, \citenamefont {Brouard}, \citenamefont {Johansson}, \citenamefont
  {Shepperson},\ and\ \citenamefont {Stapelfeldt}}]{Christensen2014}%
  \BibitemOpen
  \bibfield  {author} {\bibinfo {author} {\bibfnamefont {L.}~\bibnamefont
  {Christensen}}, \bibinfo {author} {\bibfnamefont {J.~H.}\ \bibnamefont
  {Nielsen}}, \bibinfo {author} {\bibfnamefont {C.~B.}\ \bibnamefont {Brandt}},
  \bibinfo {author} {\bibfnamefont {C.~B.}\ \bibnamefont {Madsen}}, \bibinfo
  {author} {\bibfnamefont {L.~B.}\ \bibnamefont {Madsen}}, \bibinfo {author}
  {\bibfnamefont {C.~S.}\ \bibnamefont {Slater}}, \bibinfo {author}
  {\bibfnamefont {A.}~\bibnamefont {Lauer}}, \bibinfo {author} {\bibfnamefont
  {M.}~\bibnamefont {Brouard}}, \bibinfo {author} {\bibfnamefont {M.~P.}\
  \bibnamefont {Johansson}}, \bibinfo {author} {\bibfnamefont {B.}~\bibnamefont
  {Shepperson}}, \ and\ \bibinfo {author} {\bibfnamefont {H.}~\bibnamefont
  {Stapelfeldt}},\ }\href {\doibase 10.1103/PhysRevLett.113.073005} {\bibfield
  {journal} {\bibinfo  {journal} {Phys. Rev. Lett.}\ }\textbf {\bibinfo
  {volume} {113}},\ \bibinfo {pages} {073005} (\bibinfo {year}
  {2014})}\BibitemShut {NoStop}%
\bibitem [{\citenamefont {Erk}\ \emph {et~al.}(2014)\citenamefont {Erk},
  \citenamefont {Boll}, \citenamefont {Trippel}, \citenamefont {Anielski},
  \citenamefont {Foucar}, \citenamefont {Rudek}, \citenamefont {Epp},
  \citenamefont {Coffee}, \citenamefont {Carron}, \citenamefont {Schorb},
  \citenamefont {Ferguson}, \citenamefont {Swiggers}, \citenamefont {Bozek},
  \citenamefont {Simon}, \citenamefont {Marchenko}, \citenamefont
  {K{\"{u}}pper}, \citenamefont {Schlichting}, \citenamefont {Ullrich},
  \citenamefont {Bostedt}, \citenamefont {Rolles},\ and\ \citenamefont
  {Rudenko}}]{Erk2014}%
  \BibitemOpen
  \bibfield  {author} {\bibinfo {author} {\bibfnamefont {B.}~\bibnamefont
  {Erk}}, \bibinfo {author} {\bibfnamefont {R.}~\bibnamefont {Boll}}, \bibinfo
  {author} {\bibfnamefont {S.}~\bibnamefont {Trippel}}, \bibinfo {author}
  {\bibfnamefont {D.}~\bibnamefont {Anielski}}, \bibinfo {author}
  {\bibfnamefont {L.}~\bibnamefont {Foucar}}, \bibinfo {author} {\bibfnamefont
  {B.}~\bibnamefont {Rudek}}, \bibinfo {author} {\bibfnamefont {S.~W.}\
  \bibnamefont {Epp}}, \bibinfo {author} {\bibfnamefont {R.}~\bibnamefont
  {Coffee}}, \bibinfo {author} {\bibfnamefont {S.}~\bibnamefont {Carron}},
  \bibinfo {author} {\bibfnamefont {S.}~\bibnamefont {Schorb}}, \bibinfo
  {author} {\bibfnamefont {K.~R.}\ \bibnamefont {Ferguson}}, \bibinfo {author}
  {\bibfnamefont {M.}~\bibnamefont {Swiggers}}, \bibinfo {author}
  {\bibfnamefont {J.~D.}\ \bibnamefont {Bozek}}, \bibinfo {author}
  {\bibfnamefont {M.}~\bibnamefont {Simon}}, \bibinfo {author} {\bibfnamefont
  {T.}~\bibnamefont {Marchenko}}, \bibinfo {author} {\bibfnamefont
  {J.}~\bibnamefont {K{\"{u}}pper}}, \bibinfo {author} {\bibfnamefont
  {I.}~\bibnamefont {Schlichting}}, \bibinfo {author} {\bibfnamefont
  {J.}~\bibnamefont {Ullrich}}, \bibinfo {author} {\bibfnamefont
  {C.}~\bibnamefont {Bostedt}}, \bibinfo {author} {\bibfnamefont
  {D.}~\bibnamefont {Rolles}}, \ and\ \bibinfo {author} {\bibfnamefont
  {A.}~\bibnamefont {Rudenko}},\ }\href {\doibase 10.1126/science.1253607}
  {\bibfield  {journal} {\bibinfo  {journal} {Science}\ }\textbf {\bibinfo
  {volume} {345}},\ \bibinfo {pages} {288} (\bibinfo {year}
  {2014})}\BibitemShut {NoStop}%
\bibitem [{\citenamefont {Boll}\ \emph {et~al.}(2016)\citenamefont {Boll},
  \citenamefont {Erk}, \citenamefont {Coffee}, \citenamefont {Trippel},
  \citenamefont {Kierspel}, \citenamefont {Bomme}, \citenamefont {Bozek},
  \citenamefont {Burkett}, \citenamefont {Carron}, \citenamefont {Ferguson},
  \citenamefont {Foucar}, \citenamefont {K{\"{u}}pper}, \citenamefont
  {Marchenko}, \citenamefont {Miron}, \citenamefont {Patanen}, \citenamefont
  {Osipov}, \citenamefont {Schorb}, \citenamefont {Simon}, \citenamefont
  {Swiggers}, \citenamefont {Techert}, \citenamefont {Ueda}, \citenamefont
  {Bostedt}, \citenamefont {Rolles},\ and\ \citenamefont {Rudenko}}]{Boll2016}%
  \BibitemOpen
  \bibfield  {author} {\bibinfo {author} {\bibfnamefont {R.}~\bibnamefont
  {Boll}}, \bibinfo {author} {\bibfnamefont {B.}~\bibnamefont {Erk}}, \bibinfo
  {author} {\bibfnamefont {R.}~\bibnamefont {Coffee}}, \bibinfo {author}
  {\bibfnamefont {S.}~\bibnamefont {Trippel}}, \bibinfo {author} {\bibfnamefont
  {T.}~\bibnamefont {Kierspel}}, \bibinfo {author} {\bibfnamefont
  {C.}~\bibnamefont {Bomme}}, \bibinfo {author} {\bibfnamefont {J.~D.}\
  \bibnamefont {Bozek}}, \bibinfo {author} {\bibfnamefont {M.}~\bibnamefont
  {Burkett}}, \bibinfo {author} {\bibfnamefont {S.}~\bibnamefont {Carron}},
  \bibinfo {author} {\bibfnamefont {K.~R.}\ \bibnamefont {Ferguson}}, \bibinfo
  {author} {\bibfnamefont {L.}~\bibnamefont {Foucar}}, \bibinfo {author}
  {\bibfnamefont {J.}~\bibnamefont {K{\"{u}}pper}}, \bibinfo {author}
  {\bibfnamefont {T.}~\bibnamefont {Marchenko}}, \bibinfo {author}
  {\bibfnamefont {C.}~\bibnamefont {Miron}}, \bibinfo {author} {\bibfnamefont
  {M.}~\bibnamefont {Patanen}}, \bibinfo {author} {\bibfnamefont
  {T.}~\bibnamefont {Osipov}}, \bibinfo {author} {\bibfnamefont
  {S.}~\bibnamefont {Schorb}}, \bibinfo {author} {\bibfnamefont
  {M.}~\bibnamefont {Simon}}, \bibinfo {author} {\bibfnamefont
  {M.}~\bibnamefont {Swiggers}}, \bibinfo {author} {\bibfnamefont
  {S.}~\bibnamefont {Techert}}, \bibinfo {author} {\bibfnamefont
  {K.}~\bibnamefont {Ueda}}, \bibinfo {author} {\bibfnamefont {C.}~\bibnamefont
  {Bostedt}}, \bibinfo {author} {\bibfnamefont {D.}~\bibnamefont {Rolles}}, \
  and\ \bibinfo {author} {\bibfnamefont {A.}~\bibnamefont {Rudenko}},\ }\href
  {\doibase 10.1063/1.4944344} {\bibfield  {journal} {\bibinfo  {journal}
  {Struct. Dyn.}\ }\textbf {\bibinfo {volume} {3}},\ \bibinfo {pages} {043207}
  (\bibinfo {year} {2016})}\BibitemShut {NoStop}%
\bibitem [{\citenamefont {Lee}\ and\ \citenamefont {Bersohn}(1982)}]{Lee1982}%
  \BibitemOpen
  \bibfield  {author} {\bibinfo {author} {\bibfnamefont {S.~J.}\ \bibnamefont
  {Lee}}\ and\ \bibinfo {author} {\bibfnamefont {R.}~\bibnamefont {Bersohn}},\
  }\href {\doibase Doi 10.1021/J100394a028} {\bibfield  {journal} {\bibinfo
  {journal} {J. Phys. Chem.}\ }\textbf {\bibinfo {volume} {86}},\ \bibinfo
  {pages} {728} (\bibinfo {year} {1982})}\BibitemShut {NoStop}%
\bibitem [{\citenamefont {Butler}\ \emph {et~al.}(1987)\citenamefont {Butler},
  \citenamefont {Hintsa}, \citenamefont {Shane},\ and\ \citenamefont
  {Lee}}]{Butler1987}%
  \BibitemOpen
  \bibfield  {author} {\bibinfo {author} {\bibfnamefont {L.~J.}\ \bibnamefont
  {Butler}}, \bibinfo {author} {\bibfnamefont {E.~J.}\ \bibnamefont {Hintsa}},
  \bibinfo {author} {\bibfnamefont {S.~F.}\ \bibnamefont {Shane}}, \ and\
  \bibinfo {author} {\bibfnamefont {Y.~T.}\ \bibnamefont {Lee}},\ }\href
  {\doibase 10.1063/1.452155} {\bibfield  {journal} {\bibinfo  {journal} {J.
  Chem. Phys.}\ }\textbf {\bibinfo {volume} {86}},\ \bibinfo {pages} {2051}
  (\bibinfo {year} {1987})}\BibitemShut {NoStop}%
\bibitem [{\citenamefont {Xu}\ \emph {et~al.}(2002)\citenamefont {Xu},
  \citenamefont {Guo}, \citenamefont {Liu}, \citenamefont {Ma}, \citenamefont
  {Dai},\ and\ \citenamefont {Sha}}]{Xu2002}%
  \BibitemOpen
  \bibfield  {author} {\bibinfo {author} {\bibfnamefont {H.}~\bibnamefont
  {Xu}}, \bibinfo {author} {\bibfnamefont {Y.}~\bibnamefont {Guo}}, \bibinfo
  {author} {\bibfnamefont {S.}~\bibnamefont {Liu}}, \bibinfo {author}
  {\bibfnamefont {X.}~\bibnamefont {Ma}}, \bibinfo {author} {\bibfnamefont
  {D.}~\bibnamefont {Dai}}, \ and\ \bibinfo {author} {\bibfnamefont
  {G.}~\bibnamefont {Sha}},\ }\href {\doibase 10.1063/1.1503316} {\bibfield
  {journal} {\bibinfo  {journal} {J. Chem. Phys.}\ }\textbf {\bibinfo {volume}
  {117}},\ \bibinfo {pages} {5722} (\bibinfo {year} {2002})}\BibitemShut
  {NoStop}%
\bibitem [{\citenamefont {Schmitt}\ and\ \citenamefont
  {Comes}(1987)}]{Schmitt1987}%
  \BibitemOpen
  \bibfield  {author} {\bibinfo {author} {\bibfnamefont {G.}~\bibnamefont
  {Schmitt}}\ and\ \bibinfo {author} {\bibfnamefont {F.~J.}\ \bibnamefont
  {Comes}},\ }\href {\doibase 10.1016/1010-6030(87)80002-3} {\bibfield
  {journal} {\bibinfo  {journal} {J. Photochem. Photobiol. A}\ }\textbf
  {\bibinfo {volume} {41}},\ \bibinfo {pages} {13} (\bibinfo {year}
  {1987})}\BibitemShut {NoStop}%
\bibitem [{\citenamefont {Senapati}\ \emph {et~al.}(2002)\citenamefont
  {Senapati}, \citenamefont {Kavita},\ and\ \citenamefont
  {Das}}]{Senapati2002}%
  \BibitemOpen
  \bibfield  {author} {\bibinfo {author} {\bibfnamefont {D.}~\bibnamefont
  {Senapati}}, \bibinfo {author} {\bibfnamefont {K.}~\bibnamefont {Kavita}}, \
  and\ \bibinfo {author} {\bibfnamefont {P.~K.}\ \bibnamefont {Das}},\
  }\href@noop {} {\bibfield  {journal} {\bibinfo  {journal} {J. Phys. Chem. A}\
  }\textbf {\bibinfo {volume} {106}},\ \bibinfo {pages} {8479} (\bibinfo {year}
  {2002})}\BibitemShut {NoStop}%
\bibitem [{\citenamefont {Senapati}\ and\ \citenamefont
  {Das}(2004)}]{Senapati2004}%
  \BibitemOpen
  \bibfield  {author} {\bibinfo {author} {\bibfnamefont {D.}~\bibnamefont
  {Senapati}}\ and\ \bibinfo {author} {\bibfnamefont {P.~K.}\ \bibnamefont
  {Das}},\ }\href {\doibase 10.1016/j.cplett.2004.06.095} {\bibfield  {journal}
  {\bibinfo  {journal} {Chem. Phys. Lett.}\ }\textbf {\bibinfo {volume}
  {393}},\ \bibinfo {pages} {535} (\bibinfo {year} {2004})}\BibitemShut
  {NoStop}%
\bibitem [{\citenamefont {Zhang}\ \emph {et~al.}(2005)\citenamefont {Zhang},
  \citenamefont {Ng}, \citenamefont {Qi}, \citenamefont {Lam},\ and\
  \citenamefont {Li}}]{Zhang2005}%
  \BibitemOpen
  \bibfield  {author} {\bibinfo {author} {\bibfnamefont {T.}~\bibnamefont
  {Zhang}}, \bibinfo {author} {\bibfnamefont {C.~Y.}\ \bibnamefont {Ng}},
  \bibinfo {author} {\bibfnamefont {F.}~\bibnamefont {Qi}}, \bibinfo {author}
  {\bibfnamefont {C.-S.}\ \bibnamefont {Lam}}, \ and\ \bibinfo {author}
  {\bibfnamefont {W.-K.}\ \bibnamefont {Li}},\ }\href {\doibase
  10.1063/1.2074507} {\bibfield  {journal} {\bibinfo  {journal} {J. Chem.
  Phys.}\ }\textbf {\bibinfo {volume} {123}},\ \bibinfo {pages} {174316}
  (\bibinfo {year} {2005})}\BibitemShut {NoStop}%
\bibitem [{\citenamefont {Cheng}\ \emph {et~al.}(2016)\citenamefont {Cheng},
  \citenamefont {Lin}, \citenamefont {Hu}, \citenamefont {Du},\ and\
  \citenamefont {Zhu}}]{Cheng2016}%
  \BibitemOpen
  \bibfield  {author} {\bibinfo {author} {\bibfnamefont {M.}~\bibnamefont
  {Cheng}}, \bibinfo {author} {\bibfnamefont {D.}~\bibnamefont {Lin}}, \bibinfo
  {author} {\bibfnamefont {L.}~\bibnamefont {Hu}}, \bibinfo {author}
  {\bibfnamefont {Y.}~\bibnamefont {Du}}, \ and\ \bibinfo {author}
  {\bibfnamefont {Q.}~\bibnamefont {Zhu}},\ }\href {\doibase
  10.1039/C5CP06080J} {\bibfield  {journal} {\bibinfo  {journal} {Phys. Chem.
  Chem. Phys.}\ }\textbf {\bibinfo {volume} {18}},\ \bibinfo {pages} {3165}
  (\bibinfo {year} {2016})}\BibitemShut {NoStop}%
\bibitem [{\citenamefont {El-Khoury}\ \emph {et~al.}(2010)\citenamefont
  {El-Khoury}, \citenamefont {Pal}, \citenamefont {Mereshchenko},\ and\
  \citenamefont {Tarnovsky}}]{El-Khoury2010}%
  \BibitemOpen
  \bibfield  {author} {\bibinfo {author} {\bibfnamefont {P.~Z.}\ \bibnamefont
  {El-Khoury}}, \bibinfo {author} {\bibfnamefont {S.~K.}\ \bibnamefont {Pal}},
  \bibinfo {author} {\bibfnamefont {A.~S.}\ \bibnamefont {Mereshchenko}}, \
  and\ \bibinfo {author} {\bibfnamefont {A.~N.}\ \bibnamefont {Tarnovsky}},\
  }\href {\doibase 10.1016/j.cplett.2010.04.065} {\bibfield  {journal}
  {\bibinfo  {journal} {Chem. Phys. Lett.}\ }\textbf {\bibinfo {volume}
  {493}},\ \bibinfo {pages} {61} (\bibinfo {year} {2010})}\BibitemShut
  {NoStop}%
\bibitem [{\citenamefont {Tang}\ \emph {et~al.}(2010)\citenamefont {Tang},
  \citenamefont {Peng}, \citenamefont {Spears},\ and\ \citenamefont
  {Sension}}]{Tang2010}%
  \BibitemOpen
  \bibfield  {author} {\bibinfo {author} {\bibfnamefont {K.-C.}\ \bibnamefont
  {Tang}}, \bibinfo {author} {\bibfnamefont {J.}~\bibnamefont {Peng}}, \bibinfo
  {author} {\bibfnamefont {K.~G.}\ \bibnamefont {Spears}}, \ and\ \bibinfo
  {author} {\bibfnamefont {R.~J.}\ \bibnamefont {Sension}},\ }\href {\doibase
  10.1063/1.3374680} {\bibfield  {journal} {\bibinfo  {journal} {J. Chem.
  Phys.}\ }\textbf {\bibinfo {volume} {132}},\ \bibinfo {pages} {141102}
  (\bibinfo {year} {2010})}\BibitemShut {NoStop}%
\bibitem [{\citenamefont {Rosenstock}\ \emph {et~al.}(1977)\citenamefont
  {Rosenstock}, \citenamefont {Draxl}, \citenamefont {Steiner},\ and\
  \citenamefont {Herron}}]{Rosenstock1977}%
  \BibitemOpen
  \bibfield  {author} {\bibinfo {author} {\bibfnamefont {H.~M.}\ \bibnamefont
  {Rosenstock}}, \bibinfo {author} {\bibfnamefont {K.}~\bibnamefont {Draxl}},
  \bibinfo {author} {\bibfnamefont {B.~W.}\ \bibnamefont {Steiner}}, \ and\
  \bibinfo {author} {\bibfnamefont {J.~T.}\ \bibnamefont {Herron}},\
  }\href@noop {} {\bibfield  {journal} {\bibinfo  {journal} {J. Phys. Chem.
  Ref. Data}\ }\textbf {\bibinfo {volume} {6}},\ \bibinfo {pages} {Supplement
  No. 1} (\bibinfo {year} {1977})}\BibitemShut {NoStop}%
\bibitem [{\citenamefont {Kudchadker}\ and\ \citenamefont
  {Kudchadker}(1978)}]{Kudchadker1978}%
  \BibitemOpen
  \bibfield  {author} {\bibinfo {author} {\bibfnamefont {S.~A.}\ \bibnamefont
  {Kudchadker}}\ and\ \bibinfo {author} {\bibfnamefont {A.~P.}\ \bibnamefont
  {Kudchadker}},\ }\href {\doibase 10.1063/1.555585} {\bibfield  {journal}
  {\bibinfo  {journal} {J. Phys. Chem. Ref. Data}\ }\textbf {\bibinfo {volume}
  {7}},\ \bibinfo {pages} {1285} (\bibinfo {year} {1978})}\BibitemShut
  {NoStop}%
\bibitem [{\citenamefont {Str{\"{u}}der}\ \emph {et~al.}(2010)\citenamefont
  {Str{\"{u}}der}, \citenamefont {Epp}, \citenamefont {Rolles}, \citenamefont
  {Hartmann}, \citenamefont {Holl}, \citenamefont {Lutz}, \citenamefont
  {Soltau}, \citenamefont {Eckart}, \citenamefont {Reich}, \citenamefont
  {Heinzinger}, \citenamefont {Thamm}, \citenamefont {Rudenko}, \citenamefont
  {Krasniqi}, \citenamefont {K{\"{u}}hnel}, \citenamefont {Bauer},
  \citenamefont {Schr{\"{o}}ter}, \citenamefont {Moshammer}, \citenamefont
  {Techert}, \citenamefont {Miessner}, \citenamefont {Porro}, \citenamefont
  {H{\"{a}}lker}, \citenamefont {Meidinger}, \citenamefont {Kimmel},
  \citenamefont {Andritschke}, \citenamefont {Schopper}, \citenamefont
  {Weidenspointner}, \citenamefont {Ziegler}, \citenamefont {Pietschner},
  \citenamefont {Herrmann}, \citenamefont {Pietsch}, \citenamefont {Walenta},
  \citenamefont {Leitenberger}, \citenamefont {Bostedt}, \citenamefont
  {M{\"{o}}ller}, \citenamefont {Rupp}, \citenamefont {Adolph}, \citenamefont
  {Graafsma}, \citenamefont {Hirsemann}, \citenamefont {G{\"{a}}rtner},
  \citenamefont {Richter}, \citenamefont {Foucar}, \citenamefont {Shoeman},
  \citenamefont {Schlichting},\ and\ \citenamefont {Ullrich}}]{Struder2010}%
  \BibitemOpen
  \bibfield  {author} {\bibinfo {author} {\bibfnamefont {L.}~\bibnamefont
  {Str{\"{u}}der}}, \bibinfo {author} {\bibfnamefont {S.}~\bibnamefont {Epp}},
  \bibinfo {author} {\bibfnamefont {D.}~\bibnamefont {Rolles}}, \bibinfo
  {author} {\bibfnamefont {R.}~\bibnamefont {Hartmann}}, \bibinfo {author}
  {\bibfnamefont {P.}~\bibnamefont {Holl}}, \bibinfo {author} {\bibfnamefont
  {G.}~\bibnamefont {Lutz}}, \bibinfo {author} {\bibfnamefont {H.}~\bibnamefont
  {Soltau}}, \bibinfo {author} {\bibfnamefont {R.}~\bibnamefont {Eckart}},
  \bibinfo {author} {\bibfnamefont {C.}~\bibnamefont {Reich}}, \bibinfo
  {author} {\bibfnamefont {K.}~\bibnamefont {Heinzinger}}, \bibinfo {author}
  {\bibfnamefont {C.}~\bibnamefont {Thamm}}, \bibinfo {author} {\bibfnamefont
  {A.}~\bibnamefont {Rudenko}}, \bibinfo {author} {\bibfnamefont
  {F.}~\bibnamefont {Krasniqi}}, \bibinfo {author} {\bibfnamefont {K.-U.}\
  \bibnamefont {K{\"{u}}hnel}}, \bibinfo {author} {\bibfnamefont
  {C.}~\bibnamefont {Bauer}}, \bibinfo {author} {\bibfnamefont {C.-D.}\
  \bibnamefont {Schr{\"{o}}ter}}, \bibinfo {author} {\bibfnamefont
  {R.}~\bibnamefont {Moshammer}}, \bibinfo {author} {\bibfnamefont
  {S.}~\bibnamefont {Techert}}, \bibinfo {author} {\bibfnamefont
  {D.}~\bibnamefont {Miessner}}, \bibinfo {author} {\bibfnamefont
  {M.}~\bibnamefont {Porro}}, \bibinfo {author} {\bibfnamefont
  {O.}~\bibnamefont {H{\"{a}}lker}}, \bibinfo {author} {\bibfnamefont
  {N.}~\bibnamefont {Meidinger}}, \bibinfo {author} {\bibfnamefont
  {N.}~\bibnamefont {Kimmel}}, \bibinfo {author} {\bibfnamefont
  {R.}~\bibnamefont {Andritschke}}, \bibinfo {author} {\bibfnamefont
  {F.}~\bibnamefont {Schopper}}, \bibinfo {author} {\bibfnamefont
  {G.}~\bibnamefont {Weidenspointner}}, \bibinfo {author} {\bibfnamefont
  {A.}~\bibnamefont {Ziegler}}, \bibinfo {author} {\bibfnamefont
  {D.}~\bibnamefont {Pietschner}}, \bibinfo {author} {\bibfnamefont
  {S.}~\bibnamefont {Herrmann}}, \bibinfo {author} {\bibfnamefont
  {U.}~\bibnamefont {Pietsch}}, \bibinfo {author} {\bibfnamefont
  {A.}~\bibnamefont {Walenta}}, \bibinfo {author} {\bibfnamefont
  {W.}~\bibnamefont {Leitenberger}}, \bibinfo {author} {\bibfnamefont
  {C.}~\bibnamefont {Bostedt}}, \bibinfo {author} {\bibfnamefont
  {T.}~\bibnamefont {M{\"{o}}ller}}, \bibinfo {author} {\bibfnamefont
  {D.}~\bibnamefont {Rupp}}, \bibinfo {author} {\bibfnamefont {M.}~\bibnamefont
  {Adolph}}, \bibinfo {author} {\bibfnamefont {H.}~\bibnamefont {Graafsma}},
  \bibinfo {author} {\bibfnamefont {H.}~\bibnamefont {Hirsemann}}, \bibinfo
  {author} {\bibfnamefont {K.}~\bibnamefont {G{\"{a}}rtner}}, \bibinfo {author}
  {\bibfnamefont {R.}~\bibnamefont {Richter}}, \bibinfo {author} {\bibfnamefont
  {L.}~\bibnamefont {Foucar}}, \bibinfo {author} {\bibfnamefont {R.~L.}\
  \bibnamefont {Shoeman}}, \bibinfo {author} {\bibfnamefont {I.}~\bibnamefont
  {Schlichting}}, \ and\ \bibinfo {author} {\bibfnamefont {J.}~\bibnamefont
  {Ullrich}},\ }\href {\doibase 10.1016/j.nima.2009.12.053} {\bibfield
  {journal} {\bibinfo  {journal} {Nucl. Instr. Meth. Phys. Res. A}\ }\textbf
  {\bibinfo {volume} {614}},\ \bibinfo {pages} {483} (\bibinfo {year}
  {2010})}\BibitemShut {NoStop}%
\bibitem [{\citenamefont {Erk}\ \emph {et~al.}(2017)\citenamefont {Erk},
  \citenamefont {M{\"{u}}ller}, \citenamefont {Bomme}, \citenamefont {Boll},
  \citenamefont {Brenner}, \citenamefont {Chapman}, \citenamefont {Correa},
  \citenamefont {Dachraoui}, \citenamefont {D{\"{u}}sterer}, \citenamefont
  {Dziarzhytski}, \citenamefont {Eisebitt}, \citenamefont {Feldhaus},
  \citenamefont {Graafsma}, \citenamefont {Grunewald}, \citenamefont
  {Gumprecht}, \citenamefont {Hartmann}, \citenamefont {Hauser}, \citenamefont
  {Keitel}, \citenamefont {Kuhlmann}, \citenamefont {M{\"{u}}ller},
  \citenamefont {E.}, \citenamefont {Ramm}, \citenamefont {Rupp}, \citenamefont
  {Rompotis}, \citenamefont {Sauppe}, \citenamefont {Savelyev}, \citenamefont
  {Schlichting}, \citenamefont {Str{\"{u}}der}, \citenamefont {Swiderski},
  \citenamefont {Techert}, \citenamefont {Tiedtke}, \citenamefont {Tilp},
  \citenamefont {Treusch}, \citenamefont {Ullrich}, \citenamefont {Moshammer},
  \citenamefont {M{\"{o}}ller},\ and\ \citenamefont {Rolles}}]{Erk2017}%
  \BibitemOpen
  \bibfield  {author} {\bibinfo {author} {\bibfnamefont {B.}~\bibnamefont
  {Erk}}, \bibinfo {author} {\bibfnamefont {J.~P.}\ \bibnamefont
  {M{\"{u}}ller}}, \bibinfo {author} {\bibfnamefont {C.}~\bibnamefont {Bomme}},
  \bibinfo {author} {\bibfnamefont {R.}~\bibnamefont {Boll}}, \bibinfo {author}
  {\bibfnamefont {G.}~\bibnamefont {Brenner}}, \bibinfo {author} {\bibfnamefont
  {H.~N.}\ \bibnamefont {Chapman}}, \bibinfo {author} {\bibfnamefont
  {J.}~\bibnamefont {Correa}}, \bibinfo {author} {\bibfnamefont
  {H.}~\bibnamefont {Dachraoui}}, \bibinfo {author} {\bibfnamefont
  {S.}~\bibnamefont {D{\"{u}}sterer}}, \bibinfo {author} {\bibfnamefont
  {S.}~\bibnamefont {Dziarzhytski}}, \bibinfo {author} {\bibfnamefont
  {S.}~\bibnamefont {Eisebitt}}, \bibinfo {author} {\bibfnamefont
  {J.}~\bibnamefont {Feldhaus}}, \bibinfo {author} {\bibfnamefont
  {H.}~\bibnamefont {Graafsma}}, \bibinfo {author} {\bibfnamefont
  {S.}~\bibnamefont {Grunewald}}, \bibinfo {author} {\bibfnamefont
  {L.}~\bibnamefont {Gumprecht}}, \bibinfo {author} {\bibfnamefont
  {R.}~\bibnamefont {Hartmann}}, \bibinfo {author} {\bibfnamefont
  {G.}~\bibnamefont {Hauser}}, \bibinfo {author} {\bibfnamefont
  {B.}~\bibnamefont {Keitel}}, \bibinfo {author} {\bibfnamefont
  {M.}~\bibnamefont {Kuhlmann}}, \bibinfo {author} {\bibfnamefont
  {E.}~\bibnamefont {M{\"{u}}ller}}, \bibinfo {author} {\bibfnamefont
  {P.}~\bibnamefont {E.}}, \bibinfo {author} {\bibfnamefont {D.}~\bibnamefont
  {Ramm}}, \bibinfo {author} {\bibfnamefont {D.}~\bibnamefont {Rupp}}, \bibinfo
  {author} {\bibfnamefont {D.}~\bibnamefont {Rompotis}}, \bibinfo {author}
  {\bibfnamefont {M.}~\bibnamefont {Sauppe}}, \bibinfo {author} {\bibfnamefont
  {E.}~\bibnamefont {Savelyev}}, \bibinfo {author} {\bibfnamefont
  {I.}~\bibnamefont {Schlichting}}, \bibinfo {author} {\bibfnamefont
  {L.}~\bibnamefont {Str{\"{u}}der}}, \bibinfo {author} {\bibfnamefont
  {A.}~\bibnamefont {Swiderski}}, \bibinfo {author} {\bibfnamefont
  {S.}~\bibnamefont {Techert}}, \bibinfo {author} {\bibfnamefont
  {K.}~\bibnamefont {Tiedtke}}, \bibinfo {author} {\bibfnamefont
  {T.}~\bibnamefont {Tilp}}, \bibinfo {author} {\bibfnamefont {R.}~\bibnamefont
  {Treusch}}, \bibinfo {author} {\bibfnamefont {J.}~\bibnamefont {Ullrich}},
  \bibinfo {author} {\bibfnamefont {R.}~\bibnamefont {Moshammer}}, \bibinfo
  {author} {\bibfnamefont {T.}~\bibnamefont {M{\"{o}}ller}}, \ and\ \bibinfo
  {author} {\bibfnamefont {D.}~\bibnamefont {Rolles}},\ }\href@noop {}
  {\bibfield  {journal} {\bibinfo  {journal} {in preparation}\ } (\bibinfo
  {year} {2017})}\BibitemShut {NoStop}%
\bibitem [{\citenamefont {Redlin}\ \emph {et~al.}(2011)\citenamefont {Redlin},
  \citenamefont {Al-Shemmary}, \citenamefont {Azima}, \citenamefont
  {Stojanovic}, \citenamefont {Tavella}, \citenamefont {Will},\ and\
  \citenamefont {D{\"{u}}sterer}}]{Redlin2011}%
  \BibitemOpen
  \bibfield  {author} {\bibinfo {author} {\bibfnamefont {H.}~\bibnamefont
  {Redlin}}, \bibinfo {author} {\bibfnamefont {A.}~\bibnamefont {Al-Shemmary}},
  \bibinfo {author} {\bibfnamefont {A.}~\bibnamefont {Azima}}, \bibinfo
  {author} {\bibfnamefont {N.}~\bibnamefont {Stojanovic}}, \bibinfo {author}
  {\bibfnamefont {F.}~\bibnamefont {Tavella}}, \bibinfo {author} {\bibfnamefont
  {I.}~\bibnamefont {Will}}, \ and\ \bibinfo {author} {\bibfnamefont
  {S.}~\bibnamefont {D{\"{u}}sterer}},\ }\href {\doibase
  10.1016/j.nima.2010.09.159} {\bibfield  {journal} {\bibinfo  {journal} {Nucl.
  Instr. Meth. Phys. Res. A}\ }\textbf {\bibinfo {volume} {635}},\ \bibinfo
  {pages} {S88} (\bibinfo {year} {2011})}\BibitemShut {NoStop}%
\bibitem [{\citenamefont {Eppink}\ and\ \citenamefont
  {Parker}(1997)}]{Eppink1997}%
  \BibitemOpen
  \bibfield  {author} {\bibinfo {author} {\bibfnamefont {A.~T. J.~B.}\
  \bibnamefont {Eppink}}\ and\ \bibinfo {author} {\bibfnamefont {D.~H.}\
  \bibnamefont {Parker}},\ }\href {\doibase 10.1063/1.1148310} {\bibfield
  {journal} {\bibinfo  {journal} {Rev. Sci. Instrum.}\ }\textbf {\bibinfo
  {volume} {68}},\ \bibinfo {pages} {3477} (\bibinfo {year}
  {1997})}\BibitemShut {NoStop}%
\bibitem [{\citenamefont {Clark}\ \emph {et~al.}(2012)\citenamefont {Clark},
  \citenamefont {Crooks}, \citenamefont {Sedgwick}, \citenamefont {Turchetta},
  \citenamefont {Lee}, \citenamefont {John}, \citenamefont {Wilman},
  \citenamefont {Hill}, \citenamefont {Halford}, \citenamefont {Slater},
  \citenamefont {Winter}, \citenamefont {Yuen}, \citenamefont {Gardiner},
  \citenamefont {Lipciuc}, \citenamefont {Brouard}, \citenamefont
  {Nomerotski},\ and\ \citenamefont {Vallance}}]{Clark2012}%
  \BibitemOpen
  \bibfield  {author} {\bibinfo {author} {\bibfnamefont {A.~T.}\ \bibnamefont
  {Clark}}, \bibinfo {author} {\bibfnamefont {J.~P.}\ \bibnamefont {Crooks}},
  \bibinfo {author} {\bibfnamefont {I.}~\bibnamefont {Sedgwick}}, \bibinfo
  {author} {\bibfnamefont {R.}~\bibnamefont {Turchetta}}, \bibinfo {author}
  {\bibfnamefont {J.~W.~L.}\ \bibnamefont {Lee}}, \bibinfo {author}
  {\bibfnamefont {J.~J.}\ \bibnamefont {John}}, \bibinfo {author}
  {\bibfnamefont {E.~S.}\ \bibnamefont {Wilman}}, \bibinfo {author}
  {\bibfnamefont {L.}~\bibnamefont {Hill}}, \bibinfo {author} {\bibfnamefont
  {E.}~\bibnamefont {Halford}}, \bibinfo {author} {\bibfnamefont {C.~S.}\
  \bibnamefont {Slater}}, \bibinfo {author} {\bibfnamefont {B.}~\bibnamefont
  {Winter}}, \bibinfo {author} {\bibfnamefont {W.~H.}\ \bibnamefont {Yuen}},
  \bibinfo {author} {\bibfnamefont {S.~H.}\ \bibnamefont {Gardiner}}, \bibinfo
  {author} {\bibfnamefont {M.~L.}\ \bibnamefont {Lipciuc}}, \bibinfo {author}
  {\bibfnamefont {M.}~\bibnamefont {Brouard}}, \bibinfo {author} {\bibfnamefont
  {A.}~\bibnamefont {Nomerotski}}, \ and\ \bibinfo {author} {\bibfnamefont
  {C.}~\bibnamefont {Vallance}},\ }\href {\doibase 10.1021/jp309860t}
  {\bibfield  {journal} {\bibinfo  {journal} {J. Phys. Chem. A}\ }\textbf
  {\bibinfo {volume} {116}},\ \bibinfo {pages} {10897} (\bibinfo {year}
  {2012})}\BibitemShut {NoStop}%
\bibitem [{\citenamefont {John}\ \emph {et~al.}(2012)\citenamefont {John},
  \citenamefont {Brouard}, \citenamefont {Clark}, \citenamefont {Crooks},
  \citenamefont {Halford}, \citenamefont {Hill}, \citenamefont {Lee},
  \citenamefont {Nomerotski}, \citenamefont {Pisarczyk}, \citenamefont
  {Sedgwick}, \citenamefont {Slater}, \citenamefont {Turchetta}, \citenamefont
  {Vallance}, \citenamefont {Wilman}, \citenamefont {Winter},\ and\
  \citenamefont {Yuen}}]{John2012}%
  \BibitemOpen
  \bibfield  {author} {\bibinfo {author} {\bibfnamefont {J.~J.}\ \bibnamefont
  {John}}, \bibinfo {author} {\bibfnamefont {M.}~\bibnamefont {Brouard}},
  \bibinfo {author} {\bibfnamefont {A.}~\bibnamefont {Clark}}, \bibinfo
  {author} {\bibfnamefont {J.}~\bibnamefont {Crooks}}, \bibinfo {author}
  {\bibfnamefont {E.}~\bibnamefont {Halford}}, \bibinfo {author} {\bibfnamefont
  {L.}~\bibnamefont {Hill}}, \bibinfo {author} {\bibfnamefont {J.~W.~L.}\
  \bibnamefont {Lee}}, \bibinfo {author} {\bibfnamefont {A.}~\bibnamefont
  {Nomerotski}}, \bibinfo {author} {\bibfnamefont {R.}~\bibnamefont
  {Pisarczyk}}, \bibinfo {author} {\bibfnamefont {I.}~\bibnamefont {Sedgwick}},
  \bibinfo {author} {\bibfnamefont {C.~S.}\ \bibnamefont {Slater}}, \bibinfo
  {author} {\bibfnamefont {R.}~\bibnamefont {Turchetta}}, \bibinfo {author}
  {\bibfnamefont {C.}~\bibnamefont {Vallance}}, \bibinfo {author}
  {\bibfnamefont {E.}~\bibnamefont {Wilman}}, \bibinfo {author} {\bibfnamefont
  {B.}~\bibnamefont {Winter}}, \ and\ \bibinfo {author} {\bibfnamefont {W.~H.}\
  \bibnamefont {Yuen}},\ }\href {\doibase 10.1088/1748-0221/7/08/C08001}
  {\bibfield  {journal} {\bibinfo  {journal} {J. Instrum.}\ }\textbf {\bibinfo
  {volume} {7}},\ \bibinfo {pages} {C08001} (\bibinfo {year}
  {2012})}\BibitemShut {NoStop}%
\bibitem [{\citenamefont {Slater}\ \emph {et~al.}(2015)\citenamefont {Slater},
  \citenamefont {Blake}, \citenamefont {Brouard}, \citenamefont {Lauer},
  \citenamefont {Vallance}, \citenamefont {Bohun}, \citenamefont {Christensen},
  \citenamefont {Nielsen}, \citenamefont {Johansson},\ and\ \citenamefont
  {Stapelfeldt}}]{Slater2015}%
  \BibitemOpen
  \bibfield  {author} {\bibinfo {author} {\bibfnamefont {C.~S.}\ \bibnamefont
  {Slater}}, \bibinfo {author} {\bibfnamefont {S.}~\bibnamefont {Blake}},
  \bibinfo {author} {\bibfnamefont {M.}~\bibnamefont {Brouard}}, \bibinfo
  {author} {\bibfnamefont {A.}~\bibnamefont {Lauer}}, \bibinfo {author}
  {\bibfnamefont {C.}~\bibnamefont {Vallance}}, \bibinfo {author}
  {\bibfnamefont {C.~S.}\ \bibnamefont {Bohun}}, \bibinfo {author}
  {\bibfnamefont {L.}~\bibnamefont {Christensen}}, \bibinfo {author}
  {\bibfnamefont {J.~H.}\ \bibnamefont {Nielsen}}, \bibinfo {author}
  {\bibfnamefont {M.~P.}\ \bibnamefont {Johansson}}, \ and\ \bibinfo {author}
  {\bibfnamefont {H.}~\bibnamefont {Stapelfeldt}},\ }\href {\doibase
  10.1103/PhysRevA.91.053424} {\bibfield  {journal} {\bibinfo  {journal} {Phys.
  Rev. A}\ }\textbf {\bibinfo {volume} {91}},\ \bibinfo {pages} {053424}
  (\bibinfo {year} {2015})}\BibitemShut {NoStop}%
\bibitem [{\citenamefont {Garcia}\ \emph {et~al.}(2004)\citenamefont {Garcia},
  \citenamefont {Nahon},\ and\ \citenamefont {Powis}}]{Garcia2004}%
  \BibitemOpen
  \bibfield  {author} {\bibinfo {author} {\bibfnamefont {G.~A.}\ \bibnamefont
  {Garcia}}, \bibinfo {author} {\bibfnamefont {L.}~\bibnamefont {Nahon}}, \
  and\ \bibinfo {author} {\bibfnamefont {I.}~\bibnamefont {Powis}},\ }\href
  {\doibase 10.1063/1.1807578} {\bibfield  {journal} {\bibinfo  {journal} {Rev.
  Sci. Instrum.}\ }\textbf {\bibinfo {volume} {75}},\ \bibinfo {pages} {4989}
  (\bibinfo {year} {2004})}\BibitemShut {NoStop}%
\bibitem [{\citenamefont {Liu}\ \emph {et~al.}(2005)\citenamefont {Liu},
  \citenamefont {Zhao}, \citenamefont {Wang}, \citenamefont {Zhang},
  \citenamefont {Ma},\ and\ \citenamefont {Li}}]{Liu2005}%
  \BibitemOpen
  \bibfield  {author} {\bibinfo {author} {\bibfnamefont {K.}~\bibnamefont
  {Liu}}, \bibinfo {author} {\bibfnamefont {H.}~\bibnamefont {Zhao}}, \bibinfo
  {author} {\bibfnamefont {C.}~\bibnamefont {Wang}}, \bibinfo {author}
  {\bibfnamefont {A.}~\bibnamefont {Zhang}}, \bibinfo {author} {\bibfnamefont
  {S.}~\bibnamefont {Ma}}, \ and\ \bibinfo {author} {\bibfnamefont
  {Z.}~\bibnamefont {Li}},\ }\href {\doibase 10.1063/1.1835955} {\bibfield
  {journal} {\bibinfo  {journal} {J. Chem. Phys.}\ }\textbf {\bibinfo {volume}
  {122}},\ \bibinfo {pages} {044310} (\bibinfo {year} {2005})}\BibitemShut
  {NoStop}%
\bibitem [{\citenamefont {Kim}\ \emph {et~al.}(2014)\citenamefont {Kim},
  \citenamefont {Park},\ and\ \citenamefont {Lee}}]{Kim2014}%
  \BibitemOpen
  \bibfield  {author} {\bibinfo {author} {\bibfnamefont {H.}~\bibnamefont
  {Kim}}, \bibinfo {author} {\bibfnamefont {Y.~C.}\ \bibnamefont {Park}}, \
  and\ \bibinfo {author} {\bibfnamefont {Y.~S.}\ \bibnamefont {Lee}},\ }\href
  {\doibase 10.5012/bkcs.2014.35.3.775} {\bibfield  {journal} {\bibinfo
  {journal} {Bull. Korean Chem. Soc.}\ }\textbf {\bibinfo {volume} {35}},\
  \bibinfo {pages} {775} (\bibinfo {year} {2014})}\BibitemShut {NoStop}%
\bibitem [{\citenamefont {Bailleux}\ \emph {et~al.}(2014)\citenamefont
  {Bailleux}, \citenamefont {Duflot}, \citenamefont {Taniguchi}, \citenamefont
  {Sakai}, \citenamefont {Ozeki}, \citenamefont {Okabayashi},\ and\
  \citenamefont {Bailey}}]{Bailleux2014b}%
  \BibitemOpen
  \bibfield  {author} {\bibinfo {author} {\bibfnamefont {S.}~\bibnamefont
  {Bailleux}}, \bibinfo {author} {\bibfnamefont {D.}~\bibnamefont {Duflot}},
  \bibinfo {author} {\bibfnamefont {K.}~\bibnamefont {Taniguchi}}, \bibinfo
  {author} {\bibfnamefont {S.}~\bibnamefont {Sakai}}, \bibinfo {author}
  {\bibfnamefont {H.}~\bibnamefont {Ozeki}}, \bibinfo {author} {\bibfnamefont
  {T.}~\bibnamefont {Okabayashi}}, \ and\ \bibinfo {author} {\bibfnamefont
  {W.~C.}\ \bibnamefont {Bailey}},\ }\href {\doibase 10.1021/jp510119e}
  {\bibfield  {journal} {\bibinfo  {journal} {J. Phys. Chem. A}\ }\textbf
  {\bibinfo {volume} {118}},\ \bibinfo {pages} {11744} (\bibinfo {year}
  {2014})}\BibitemShut {NoStop}%
\bibitem [{\citenamefont {Eppink}\ and\ \citenamefont
  {Parker}(1998)}]{Parker1998}%
  \BibitemOpen
  \bibfield  {author} {\bibinfo {author} {\bibfnamefont {A.~T. J.~B.}\
  \bibnamefont {Eppink}}\ and\ \bibinfo {author} {\bibfnamefont {D.~H.}\
  \bibnamefont {Parker}},\ }\href@noop {} {\bibfield  {journal} {\bibinfo
  {journal} {J. Chem. Phys.}\ }\textbf {\bibinfo {volume} {109}},\ \bibinfo
  {pages} {4758} (\bibinfo {year} {1998})}\BibitemShut {NoStop}%
\bibitem [{\citenamefont {Attar}\ \emph {et~al.}(2014)\citenamefont {Attar},
  \citenamefont {Piticco},\ and\ \citenamefont {Leone}}]{Attar2014}%
  \BibitemOpen
  \bibfield  {author} {\bibinfo {author} {\bibfnamefont {A.~R.}\ \bibnamefont
  {Attar}}, \bibinfo {author} {\bibfnamefont {L.}~\bibnamefont {Piticco}}, \
  and\ \bibinfo {author} {\bibfnamefont {S.~R.}\ \bibnamefont {Leone}},\ }\href
  {\doibase 10.1063/1.4898375} {\bibfield  {journal} {\bibinfo  {journal} {J.
  Chem. Phys.}\ }\textbf {\bibinfo {volume} {141}},\ \bibinfo {pages} {164308}
  (\bibinfo {year} {2014})}\BibitemShut {NoStop}%
\bibitem [{\citenamefont {Andrews}\ \emph {et~al.}(1984)\citenamefont
  {Andrews}, \citenamefont {Dyke}, \citenamefont {Jonathan}, \citenamefont
  {Keddar},\ and\ \citenamefont {Morris}}]{Andrews1984}%
  \BibitemOpen
  \bibfield  {author} {\bibinfo {author} {\bibfnamefont {L.}~\bibnamefont
  {Andrews}}, \bibinfo {author} {\bibfnamefont {J.~M.}\ \bibnamefont {Dyke}},
  \bibinfo {author} {\bibfnamefont {N.}~\bibnamefont {Jonathan}}, \bibinfo
  {author} {\bibfnamefont {N.}~\bibnamefont {Keddar}}, \ and\ \bibinfo {author}
  {\bibfnamefont {A.}~\bibnamefont {Morris}},\ }\href {\doibase
  10.1021/j150654a007} {\bibfield  {journal} {\bibinfo  {journal} {J. Phys.
  Chem.}\ }\textbf {\bibinfo {volume} {88}},\ \bibinfo {pages} {1950} (\bibinfo
  {year} {1984})}\BibitemShut {NoStop}%
\bibitem [{\citenamefont {Ma}\ \emph {et~al.}(1993)\citenamefont {Ma},
  \citenamefont {Liao}, \citenamefont {Ng}, \citenamefont {Ma},\ and\
  \citenamefont {Li}}]{Ma1993a}%
  \BibitemOpen
  \bibfield  {author} {\bibinfo {author} {\bibfnamefont {Z.-X.}\ \bibnamefont
  {Ma}}, \bibinfo {author} {\bibfnamefont {C.-L.}\ \bibnamefont {Liao}},
  \bibinfo {author} {\bibfnamefont {C.~Y.}\ \bibnamefont {Ng}}, \bibinfo
  {author} {\bibfnamefont {N.~L.}\ \bibnamefont {Ma}}, \ and\ \bibinfo {author}
  {\bibfnamefont {W.-K.}\ \bibnamefont {Li}},\ }\href {\doibase
  10.1063/1.465864} {\bibfield  {journal} {\bibinfo  {journal} {J. Chem.
  Phys.}\ }\textbf {\bibinfo {volume} {99}},\ \bibinfo {pages} {6470} (\bibinfo
  {year} {1993})}\BibitemShut {NoStop}%
\end{thebibliography}
\end{document}